\documentclass[11pt]{article}

\pdfoutput=1
\usepackage[english]{babel}
\usepackage{graphicx}
\usepackage{hyperref}
\usepackage{cite}
\usepackage{subfigure}
\usepackage{amsmath}

\usepackage{rotating}        % for sidewaystable

\usepackage[top=15mm,bottom=12mm,left=30mm,right=30mm,foot=12mm,includefoot]{geometry}

\allowdisplaybreaks[2]

\newcommand{\sect}[1]{section~#1}
\newcommand{\fig}[1]{figure~#1}
\newcommand{\figs}[1]{figures~#1}
\newcommand{\tab}[1]{table~#1}
\newcommand{\eqn}[1]{equation~#1}

\newcommand{\gev}{\operatorname{GeV}}
\newcommand{\mev}{\operatorname{MeV}}

\newcommand{\ms}{\mskip 1.5mu}

\newcommand{\gsim}{\raisebox{-4pt}{%
    $\,\stackrel{\textstyle >}{\sim}\,$}}

\newcommand{\Qbar}{\overline{Q}}

\newcommand{\as}{\alpha_s}

\graphicspath{{plots/}}

\begin{document}

\begin{flushright}
DESY 19-068
\end{flushright}

\begin{center}
\vspace{4\baselineskip}
\textbf{\Large Gluons and sea quarks in the proton at low scales} \\
\vspace{3\baselineskip}
M.~Diehl and P.~Stienemeier \\
\vspace{1\baselineskip}
Deutsches Elektronen-Synchrotron DESY, Notkestr.~85, 22607 Hamburg, Germany \\
\vspace{5\baselineskip}
\parbox{0.9\textwidth}{
\textbf{Abstract:}
We study the evolution of parton distributions down to low scales by considering several of their Mellin moments.  For the initial conditions,
we use a broad array of current parton density fits.  Confirming earlier findings in the literature, we conclude that current determinations of parton distributions are incompatible with the idea that gluon or antiquark densities are generated by purely perturbative radiation as it is encoded in the DGLAP evolution equations.
}
\end{center}

\newpage

\tableofcontents

\begin{center}
\rule{0.6\textwidth}{0.3pt}
\end{center}

%%%%%%%%%%%%%%%%%%%%%%%%%%%%%%%

\section{Introduction}

An outstanding question in QCD is the relation between its basic degrees of freedom---quarks, antiquarks, and gluons---and the concept of ``constituent quarks'', which plays a major role in the description of the spectrum and static properties of hadrons (see \sect{15} of \cite{Tanabashi:2018oca} for a brief overview and references).  The internal structure of hadrons at fine spatial resolution is described by parton distribution functions (PDFs) and similar quantities such as transverse-momentum dependent distributions, generalised parton distributions, or double parton distributions.  It is natural to ask how the resulting picture of the proton as a system made of many quarks, antiquarks and gluons can be related with the picture of the proton as a bound state of just two up quarks and one down quark.

A simple and physically intuitive idea, put forward long ago, is that at coarse spatial resolution the proton contains only valence quarks, and that antiquarks and gluons (as well as quarks with low momentum fraction) are generated by radiation, i.e.\ by splitting processes like $q\to q g$ and $g \to q \bar{q}$ that can be computed in QCD perturbation theory \cite{Parisi:1976fz,Novikov:1976dd,Gluck:1977ah}.  In a technical implementation of this idea, gluon and antiquark distributions are zero at some low renormalisation scale, and the parton splitting processes encoded in the DGLAP evolution equations generate nonzero gluon and antiquark distributions at higher scales.  However, it was already found in the 1990s that PDFs constructed along these lines are in conflict with experimental data \cite{Gluck:1989ze,Gluck:1991ng,Gluck:1994uf}.  Modifying the original idea, the Dortmund group of Gl\"uck, Reya, and their collaborators performed a long series of PDF fits with a low initial renormalisation scale (typically well below $1 \gev$), so that perturbative scale evolution plays a major role in the shape of these distributions at higher scales.  This was done both in the unpolarised and in the polarised sector, see \cite{Jimenez-Delgado:2014twa,Gluck:2000dy} and references therein.

The simple scenario just described has often been used to compute parton distributions from dynamical quark models, see for instance \cite{Jaffe:1980ti} and \cite{Traini:1997jz}.  More complicated quantities can be obtained in the same manner, such as generalised parton distributions \cite{Scopetta:2002xq} or double parton distributions \cite{Rinaldi:2014ddl}.  The study in \cite{Traini:1997jz} concluded that such approaches ``tend to give a qualitative description of the data'' but are insufficient at the quantitative level.

Several approaches have been pursued to connect models at low momentum scales with PDFs.  In a formulation put forward long ago \cite{Altarelli:1973ff} and used later e.g.\ in \cite{Scopetta:1997wk,Noguera:2004jq}, the three ``constituent quarks'' of a proton have an internal structure that involves gluons and antiquarks.
A different line of work is based on the idea that virtual meson fluctuations (often referred to as ``meson cloud'') provide a natural source of antiquarks in the proton even at a low scale~\cite{Thomas:1983fh}.  We refer to  \sect{4.3.1} of \cite{Kumano:1997cy} for more detail and references, and to \cite{Traini:2013zqa,Wang:2016ndh,Kofler:2017uzq} for recent applications.  The role of gluons in this context is discussed in \cite{Koepf:1995yh,Strikman:2003gz}.  Yet a different picture emerges in the chiral quark-soliton model, where the proton has a Dirac sea of quarks and antiquarks at the typical scale $\mu \sim 600\mev$ of the model \cite{Diakonov:1996sr,Diakonov:1997vc}.  As described in these papers, gluon
distributions in this model are suppressed parametrically compared with quarks and antiquarks; in physical terms one has a scenario in which the quark and antiquark degrees of freedom of the model have themselves a structure that gives rise to gluon distributions.

The goal of the present paper is to approach the preceding discussion from a different angle.  Starting with our present knowledge of PDFs, as encoded in PDF sets fitted to experimental data by different groups, we wish to investigate the possibility that the distribution functions for gluons or antiquarks, or both, become zero (or at least small) when one evolves them backwards to low scales.  At a practical level, evolving PDFs from high to low scales is however delicate, because small changes at the starting scale of evolution quickly blow up.  We circumvent this problem by evolving backwards several Mellin moments of parton distributions.  The evolution of Mellin moments is described by simple differential equations, which can be solved numerically without numerical stability problems.  We can then check the hypothesis that a given PDF becomes zero under evolution to a low scale by checking whether several of its moments evolve to zero at one and the same scale~$\mu$.  To get a sense of whether perturbative evolution can actually be trusted at low scales, we compare the different orders for which PDF sets are available, i.e.\ leading order (LO), next-to-leading order (NLO), and next-to-next-to-leading order (NNLO).

We note that the scale evolution of PDF moments, including evolution to low scales, has been considered extensively in spin physics, with a focus on the first moment of the helicity dependent PDFs.  Indeed, the ``proton spin crisis'' was triggered by the observation that the values of these moments extracted from experiment are not consistent with the idea that at a low scale the spin $1/2$ of the proton simply arises from the helicities of its three constituent quarks.  For details we refer to the review \cite{Lampe:1998eu}, and for more recent work on the evolution of spin dependent PDF moments to \cite{Thomas:2008ga,Altenbuchinger:2010sz,deFlorian:2019egz}.  In fact, our present work is somewhat similar in spirit to the study in \cite{Altenbuchinger:2010sz}.   That work was concerned with the evolution of the helicity and the total angular momentum carried by quarks, starting at $\mu=2\gev$ with input values from lattice calculations and evolving to low scales in order to compare with quark models.  We note that significant deviations between LO and higher orders (NLO and NNLO) were found below $\mu\sim 500\mev$, and we anticipate that we will find the same in the unpolarised sector studied here.

The present work is structured as follows.  In \sect{\ref{sec:caveats}} we discuss some obvious caveats of our study.  We then introduce our notation and briefly recall the evolution of Mellin moments in \sect{\ref{sec:mellin-evo}}.  In \sect{\ref{sec:pdfs}} we present the different PDF sets that we use to compute the  starting values for evolution.  In the same section, we quantify to which parton momentum fractions $x$ one is most sensitive in a given Mellin moment, and we take a brief look at the running of the strong coupling.  In \sect{\ref{sec:results}} we investigate in detail the evolution of Mellin moments to low scales.  The main findings of our study are summarised in \sect{\ref{sec:sum}}.

\section{Caveats}
\label{sec:caveats}

Our study is subject to several caveats, which we now briefly discuss.  Perhaps the most obvious one is our use of perturbation theory down to rather low scale.  In fact, there is a substantial body of work on the behaviour of $\alpha_s$ in the low-scale regime, see for instance the review \cite{Deur:2016tte} and the recent lattice studies \cite{DallaBrida:2016kgh,Bruno:2017gxd,Zafeiropoulos:2019flq}.  However, to the best of our knowledge, there are no corresponding studies for the scale dependence of PDFs or their Mellin moments.  To assume that PDFs evolve as given by perturbation theory in a region where the running of $\alpha_s(\mu)$ is strongly affected by non-perturbative effects would in our view require some motivation.  In the present work, we use perturbation theory to evolve both the strong coupling and Mellin moments to low scales with the aim to see what one obtains in such a scenario, without strong claims that this is a valid approximation at a given scale $\mu$.  In doing so, we follow the procedure adopted in many works that connect quark models at low scale with PDFs (see the papers cited in the introduction).  By comparing perturbative results at three different orders, we will in fact get some indication of how stable the perturbative expansion is at a given scale for the quantities we are interested in.

Another caveat concerns the renormalisation scheme used to define PDFs and $\alpha_s$.  We will use the $\overline{\text{MS}}$ scheme throughout this work.  This choice is dictated by practical considerations, as it is in this scheme that the DGLAP splitting functions are available up to NNLO.  The strength of this scheme is its suitability for higher order perturbative computations, but as a downside it does not readily offer an intuitive interpretation of renormalised quantities.  One might think of other schemes for defining PDFs, e.g.\ the DIS scheme, but this will not be pursued in the present work.  (Notice that, while quark and antiquark densities have a rather straightforward physical meaning in that scheme, the same is not true for the gluon distribution).    We note that PDFs and the running coupling at LO play a special role in this context, since they are the same in a large class of renormalisation schemes.  Returning to the discussion in the previous paragraph, one might ideally want to define a scheme that allows for a physical interpretation of PDFs and that can be used in a non-perturbative setting, but such an endeavour is well beyond the scope of this work.

Finally, the very idea that antiquark or gluon distributions in a hadron are zero at a certain scale $\mu_{c}$ is somewhat problematic regarding their physical interpretation.  This is because these distributions will then typically become negative at scales just below $\mu_{c}$, at least in some $x$ range.  We will see this happen at the level of their Mellin moments.  The interpretation of PDFs as probability densities is then lost at scales below $\mu_{c}$.  One may mitigate this problem by considering a scale just above $\mu_c$, where antiquark or gluon densities would be nonzero but small (compared with their values at higher scales, or compared with the quark distributions at the same scale).

\section{Evolution of Mellin moments}
\label{sec:mellin-evo}

Let us briefly recall how the Mellin moments of PDFs depend on the renormalisation scale.  Throughout this work, we consider distributions for $n_f = 3$ active quark flavours and limit our attention to the flavour singlet combinations of quark and antiquark distributions.  One could extend our study to individual quark flavours, investigating for instance whether moments of the distribution $s(x,\mu) + \bar{s}(x,\mu)$ vanish at some low scale.  In such a case, strangeness in the proton would be generated by perturbative evolution.  To pursue this is, however, beyond the scope of this paper.

We define Mellin moments for flavour summed quark and antiquark distributions,
\begin{align}
  Q(j,\mu) &= \sum_{q=u,d,s} \int_0^1 \! d x\, x^{j-1}\, q(x,\mu) \,,
&
  \Qbar(j,\mu) &= \sum_{q=u,d,s} \int_0^1 \! d x\, x^{j-1}\, \bar{q}(x,\mu)
\end{align}
and for the gluon distribution,
\begin{align}
   G(j,\mu) &= \int_0^1 \! d x\, x^{j-1}\, g(x,\mu) \,,
\end{align}
recalling that all quantities are renormalised in the $\overline{\text{MS}}$ scheme.  The valence combination $Q - \Qbar$ evolves as
\begin{align}
\label{nons-evo}
  \frac{d}{d\log \mu^2}\, \bigl[ Q(j,\mu) - \Qbar(j,\mu) \bigr]
   &= \gamma_{\text{ns}}(j,\mu)\, \bigl[ Q(j,\mu) - \Qbar(j,\mu) \bigr] \,,
\end{align}
where the non-singlet anomalous dimension $\gamma_{\text{ns}}(j,\mu)$ depends on $\mu$ via the scale of $\as$.  In the singlet sector, we have the matrix equation
\begin{align}
  \label{sing-evo}
  \frac{d}{d\log \mu^2}\,
  \begin{pmatrix}  Q(j,\mu)
  + \Qbar(j,\mu) \\[0.15em] G(j,\mu) \end{pmatrix}
  &= \begin{pmatrix}
         \gamma_{q q}(j,\mu) & \gamma_{q g}(j,\mu) \\[0.15em]
         \gamma_{g q}(j,\mu) & \gamma_{g g}(j,\mu)
     \end{pmatrix} \,
     \begin{pmatrix}  Q(j,\mu) + \Qbar(j,\mu) \\[0.15em] G(j,\mu) \end{pmatrix} \,.
\end{align}
The anomalous dimensions have a perturbative expansion, which reads
\begin{align}
\label{gamma-expand}
  \gamma_{i}^{}(j,\mu) &=
  \sum_{k=0}^{\infty} \biggl( \frac{\as(\mu)}{2\pi} \biggr)^{k+1}\,
  \gamma_{i}^{(k)}(j) \,,
& \text{ with $i \in \{ \text{ns}, q q, q g, g q, g g \}$.}
\end{align}
Using the renormalisation group equation $d\as /d \log\mu^2 = \beta(\as)$, we rewrite \eqref{nons-evo} and \eqref{sing-evo} as evolution equations in the variable $\as$, which gives
\begin{align}
\label{nons-evo-alp}
  \frac{d}{d\as}\, \bigl[ Q(j,\as) - \Qbar(j,\as) \bigr]
   &= \frac{\gamma_{\text{ns}}(j,\as)}{\beta(\as)}\,
   \bigl[ Q(j,\as) - \Qbar(j,\as) \bigr]
\end{align}
and an analogous equation for the singlet sector.  We use the perturbative expansion of $\beta(\as)$ at the same order as for the anomalous dimensions and solve the equations numerically using the classical four-step Runge-Kutta method.  For brevity, we suppress the dependence of the Mellin moments on $\mu$ (or on the corresponding value of $\as$) henceforth.

The non-singlet equation \eqref{nons-evo-alp} can of course be solved analytically by straightforward integration.  The same holds for the singlet equation if $j=2$, in which case $G(2)$ can be eliminated by using the momentum sum rule $Q(2) + \Qbar(2) + G(2) = 1$.  We use these analytic solutions as cross checks of the numerical ones.

At odd values of $j$, the combination $Q(j) - \Qbar(j)$ of moments is connected with the proton matrix elements of local twist-two operators, and the same holds for $Q(j) + \Qbar(j)$ and for $G(j)$ if $j$ is even.  In our study, we are however interested in the antiquark moments $\Qbar(j)$ by themselves, which are not connected with local twist-two operators at any value of $j$.  Furthermore, for reasons discussed in \sect{\ref{sec:moments-x}}, we will also consider non-integer values of $j$.  We obtain the anomalous dimensions needed in \eqref{nons-evo} and \eqref{sing-evo} by computing the relevant Mellin moments of the DGLAP splitting functions.  For the splitting functions up to NLO ($k=1$) we use the exact expressions given in \cite{Ellis:1991qj}, whereas for the NNLO $(k=2)$ splitting functions we take the parameterisations given in \cite{Moch:2004pa,Vogt:2004mw}, which approximate the exact kernels and have a much simpler functional form.  In both cases, the Mellin moments are easy to compute.  We cross checked our results for the $j=2$ anomalous dimensions in the singlet sector against the analytic expressions given in  \cite{Larin:1996wd}.  For the NNLO coefficients, we also verified the constraints
\begin{align}
\gamma_{\text{ns}}^{(2)}(1) &= 0 \,,
&
\gamma_{q q}^{(2)}(2) + \gamma_{g q}^{(2)}(2) &= 0 \,,
&
\gamma_{q g}^{(2)}(2) + \gamma_{g g}^{(2)}(2) &= 0
\end{align}
from fermion number and momentum conservation for our numerical results
and find them to be satisfied within better than $2 \times 10^{-3}$ for $n_F = 3$.  We take this as evidence that the approximate forms of the NNLO splitting functions in \cite{Moch:2004pa,Vogt:2004mw} are sufficiently accurate for our purposes.  In \tab{\ref{tab:gammas}} we give the numerical values of the perturbative expansion coefficients in \eqref{gamma-expand} as functions of $n_f$.

We note that approximate expressions for the DGLAP splitting functions in the non-singlet sector are available at N$^3$LO \cite{Moch:2017uml}.  Furthermore, the N$^3$LO coefficients $\smash{\gamma_i^{(3)}(j)}$ in the singlet sector have been given in \cite{Vogt:2018miu} for $j=2$ and $n_F = 4$.  This is however not sufficient for extending the expansion coefficients given in \tab{\ref{tab:gammas}} to the next order, $k=3$, and we limit our present analysis to $k \le 2$.

\begin{table*}
\renewcommand{\arraystretch}{1.22}
\begin{center}
  \begin{tabular}{| c | l | l | l |}
      \multicolumn{4}{l}{$\gamma^{(k)}_\text{ns}(j)$} \\[0.1em]
      \hline
        & \multicolumn{3}{|c|}{$k$} \\
      \cline{2-4}
      $j$ & \multicolumn{1}{c|}{$0$} & \multicolumn{1}{c|}{$1$}               & \multicolumn{1}{c|}{$2$}                      \\
      \hline
      $1.5$                   & $-1.0588$                & $-\phantom{0}7.6282 + 0.4513 \, n_f$   & $-\phantom{0}67.070 + \phantom{0}8.8140 \, n_f + 0.0986 \, n_f^2$  \\
      $2$                     & $-1.7778$                & $-12.0656 + 0.7901 \, n_f$             & $-106.024 + 15.7285 \, n_f + 0.1536 \, n_f^2$ \\
      $2.5$                   & $-2.3286$                & $-15.2318 + 1.0584\, n_f$              & $-133.808 + 20.7085 \, n_f + 0.1914 \, n_f^2$ \\
      $3$                     & $-2.7778$                & $-17.7212 + 1.2809 \, n_f$             & $-155.615 + 24.5589 \, n_f + 0.2203 \, n_f^2$ \\
      \hline
  \end{tabular}\\[0.75em]
  \begin{tabular}{| c | l | l | l |}
      \multicolumn{4}{l}{$\gamma^{(k)}_{q q}(j)$} \\[0.1em]
      \hline
        & \multicolumn{3}{|c|}{$k$} \\
      \cline{2-4}
      $j$ & \multicolumn{1}{c|}{$0$} & \multicolumn{1}{c|}{$1$}               & \multicolumn{1}{c|}{$2$}                      \\
      \hline
      $1.5$                   & $-1.0588$                & $-\phantom{0}7.6880 + 2.5188 \, n_f$ & $-\phantom{0}70.245 + 35.6085 \, n_f + 1.0777 \, n_f^2$  \\
      $2$                     & $-1.7778$                & $-12.0823 + 1.2840 \, n_f$           & $-107.431 + 21.9562 \, n_f + 0.5844 \, n_f^2$ \\
      $2.5$                   & $-2.3286$                & $-15.2376 + 1.2431 \, n_f$           & $-134.568 + 23.5801 \, n_f + 0.4386 \, n_f^2$ \\
      $3$                     & $-2.7778$                & $-17.7236 + 1.3667 \, n_f$           & $-156.081 + 26.2653 \, n_f + 0.3816 \, n_f^2$ \\
      \hline
  \end{tabular}\\[0.75em]
  \begin{tabular}{| c | r | r | r |}
      \multicolumn{4}{l}{$\gamma^{(k)}_{q g}(j)$} \\[0.1em]
      \hline
        & \multicolumn{3}{|c|}{$k$} \\
      \cline{2-4}
      $j$ & \multicolumn{1}{c|}{$0$} & \multicolumn{1}{c|}{$1$}               & \multicolumn{1}{c|}{$2$}                      \\
      \hline
      $1.5$                   & $0.4381\, n_f$           & $6.2109\, n_f$           & $42.1828\, n_f - 1.8462\, n_f^2$ \\
      $2$                     & $0.3333\, n_f$           & $1.8858\, n_f$           & $\phantom{0}4.7030\, n_f - 1.5140\, n_f^2$ \\
      $2.5$                   & $0.2730\, n_f$           & $0.7182\, n_f$           & $-1.0327\, n_f - 1.2458\, n_f^2$ \\
      $3$                     & $0.2333\, n_f$           & $0.1840\, n_f$           & $-2.9878\, n_f - 1.0421\, n_f^2$ \\
      \hline
  \end{tabular}\\[0.75em]
  \begin{tabular}{| c | r | r | r |}
      \multicolumn{4}{l}{$\gamma^{(k)}_{g q}(j)$} \\[0.1em]
      \hline
        & \multicolumn{3}{|c|}{$k$} \\
      \cline{2-4}
      $j$ & \multicolumn{1}{c|}{$0$} & \multicolumn{1}{c|}{$1$}               & \multicolumn{1}{c|}{$2$}                      \\
      \hline
      $1.5$                   & $4.0889$                 & $21.6219 - 3.6367\, n_f$           & $120.262 - 58.7094\, n_f - 1.0024\, n_f^2$  \\
      $2$                     & $1.7778$                 & $12.0823 - 1.2840\, n_f$           & $107.431 - 21.9562\, n_f - 0.5844\, n_f^2$ \\
      $2.5$                   & $1.0920$                 & $\phantom{0}8.2191 - 0.6455\, n_f$ & $\phantom{0}76.642  - 12.7228\, n_f - 0.4180\, n_f^2$ \\
      $3$                     & $0.7778$                 & $\phantom{0}6.1566 - 0.3765\, n_f$ & $\phantom{0}58.582 - \phantom{0}8.6419\, n_f - 0.3259\, n_f^2$ \\
      \hline
  \end{tabular}\\[0.75em]
  \begin{tabular}{| c | r | r | r |}
      \multicolumn{4}{l}{$\gamma^{(k)}_{g g}(j)$} \\[0.1em]
      \hline
        & \multicolumn{3}{|c|}{$k$} \\
      \cline{2-4}
      $j$ & \multicolumn{1}{c|}{$0$} & \multicolumn{1}{c|}{$1$}               & \multicolumn{1}{c|}{$2$}                      \\
      \hline
      $1.5$                   & $6.5035 -0.3333 \, n_f$  & $\phantom{0}\,22.2636 - 8.6903\, n_f$  & $\phantom{0}62.257 - 72.4896\, n_f + 1.8677\, n_f^2$  \\
      $2$                     & $-\,0.3333 \, n_f$       & $\phantom{000.0000}\, - 1.8858\, n_f$  & $\phantom{00.0000}\, - \phantom{0}4.7029\, n_f + 1.5141\, n_f^2$        \\
      $2.5$                   & $-2.6013 -0.3333 \, n_f$ & $-12.4630 + 0.5230\, n_f$              & $-\phantom{0}98.005 + 18.6109\, n_f + 1.3748\, n_f^2$ \\
      $3$                     & $-4.2000-0.3333 \, n_f$  & $-20.8255 + 1.8512\, n_f$              & $-165.957 + 32.5239\, n_f + 1.3219\, n_f^2$ \\
      \hline
  \end{tabular}
\end{center}
\caption{\label{tab:gammas} Coefficients in the expansion \protect\eqref{gamma-expand} of anomalous dimensions at N$^{k}$LO for the moment indices $j$ considered in this work. Details on their computation are given in the text.}
\end{table*}

\section{Parton densities, their moments and the running coupling}
\label{sec:pdfs}

We perform our study with a wide range of current PDF sets, which should reflect the current knowledge and uncertainties of unpolarised parton densities.\footnote{%
Updates to some of the PDF sets used here have been presented at the DIS
2019 Workshop, see \url{https://indico.cern.ch/event/749003}.}
In table \ref{tab:pdfs} we list these sets together with some of their characteristic parameters.  In the first column, we give the full name of a set in the LHAPDF library \cite{Buckley:2014ana}, from which we take the numerical values of all PDFs. The second column shows the ``short names'' that will be used to refer to a given set throughout this paper.
A number of comments are in order.
\begin{itemize}
\item The 2018 Review of Particle Physics \cite{Tanabashi:2018oca} gives $\as(M_Z) = 0.1181 \pm 0.0011$ as world average for the strong coupling at the $Z$ mass, which corresponds to values from $0.1159$ to $0.1203$ at $2\sigma$ accuracy.  To assess the impact of the $\as$ value within a single PDF fitting approach, we take the NNLO sets of NNPDF \cite{Ball:2017nwa} for $\as(M_Z) = 0.116, 0.118, 0.120$.  We note that the value $\as(M_Z) = 0.11471$ in the ABMP set \cite{Alekhin:2017kpj} is even smaller.
\item As starting scale for evolution, we take $\mu_0 = 1.3 \gev$ for all PDF sets.  At this scale, all sets using a variable flavour number scheme have $n_f = 3$ active quarks.  The charm quark mass $m_c$ in the ABMP set is smaller than $\mu_0$, but this set uses the fixed flavour number scheme with $n_f = 3$.
\item The default PDF set in the NNPDF study \cite{Ball:2017nwa} has an intrinsic charm distribution and is hence not available for $n_f = 3$.  Evolving a $n_f = 4$ set down to low scales makes little physical sense.  We therefore take the ``perturbative charm'' variant of that study, in which charm distributions are generated by evolution, as is the case for all other PDF sets in our study.
\item In the tradition of PDF fits by the Dortmund group, the JR study \cite{Jimenez-Delgado:2014twa} takes an initial condition at very low scale, namely at $Q_0^2 = 0.8 \gev^2$.  Such an approach is of obvious interest in the context of our study.  To assess its impact on the moments of PDFs, we also include the set with a more conventional starting scale of $Q_0^2 = 2.0 \gev^2$ from the same study.  We refer to the respective sets as ``JR 08'' and ``JR 20''.
\item We include PDF sets from CJ \cite{Accardi:2016qay}, because that study pays particular attention to the region of large $x$.  We find that the PDF error bands for these PDFs, as given by the LHAPDF interface, are considerably smaller than those of any other set we studied.  We have not investigated the reasons for this and decided not to show the corresponding sets in our plots.  We will, however, include the CJ15 sets in our discussion later on.
\item The LO set of CT \cite{Dulat:2015mca}, as implemented in LHAPDF, does not give any PDF uncertainties, and we therefore exclude it from our study.
\end{itemize}

\begin{sidewaystable}[p]
\renewcommand{\arraystretch}{1.2}
\begin{center}
\begin{tabular}{|cccclccc|}
\hline
  LHAPDF identifier & short name & Ref. &
  $\as(1.3 \gev)$ & $\as(M_Z)$ &
  $m_c [\gev]$ & $m_b [\gev]$ & $x_{\text{min}}$ \\
\hline
  ABMP16\_3\_nnlo & ABMP & \protect\cite{Alekhin:2017kpj} &
    0.340 & 0.11471 & 1.252 & 3.838 & $10^{-7}$ \rule{0pt}{1.2em} \\
  CJ15lo & CJ LO & \protect\cite{Accardi:2016qay} &
    0.500 & 0.118 & 1.3\phantom{00} & 4.5\phantom{00} & $10^{-6}$ \\
  CJ15nlo & CJ NLO & &
    0.385 & 0.118 & & & \\
  CT14nlo & CT NLO & \protect\cite{Dulat:2015mca} &
    0.372 & 0.118 & 1.3\phantom{00} & 4.75\phantom{0} & $10^{-9}$ \\
  CT14nnlo & CT NNLO & &
    0.378 & 0.118 & & & \\
  HERAPDF20\_LO\_EIG & HERAPDF LO & \protect\cite{Abramowicz:2015mha} &
    0.424 & 0.130 & 1.47\phantom{0} & 4.5\phantom{00} & $9.9 \times 10^{-7}$ \\
  HERAPDF20\_NLO\_EIG & HERAPDF NLO & &
    0.374 & 0.118 & & & \\
  HERAPDF20\_NNLO\_EIG & HERAPDF NNLO & &
    0.379 & 0.118 & & & \\
  JR14NLO08FF & JR NLO 08 & \protect\cite{Jimenez-Delgado:2014twa} &
    0.347 & 0.1158 & 1.3\phantom{00} & 4.2\phantom{00} & $10^{-9}$ \\
  JR14NNLO08FF & JR NNLO 08 & &
    0.332 & 0.1136 & & & \\
  JR14NNLO20FF & JR NNLO 20 & &
    0.359 & 0.1162 & & & \\
  MMHT2014lo68cl & MMHT LO & \protect\cite{Harland-Lang:2014zoa} &
    0.482 & 0.135 & 1.4\phantom{00} & 4.75\phantom{0} & $10^{-6}$ \\
  MMHT2014nlo68cl & MMHT NLO & &
    0.397 & 0.120 & & & \\
  MMHT2014nnlo68cl & MMHT NNLO & &
    0.380 & 0.118 & & & \\
  NNPDF31\_lo\_pch\_as\_0118 & NNPDF LO & \protect\cite{Ball:2017nwa} &
    0.316 & 0.118 & 1.51\phantom{0} & 4.92\phantom{0} & $10^{-9}$ \\
  NNPDF31\_nlo\_pch\_as\_0118 & NNPDF NLO & &
    0.364 & 0.118 & & & \\
  NNPDF31\_nnlo\_pch\_as\_0118 & NNPDF NNLO 118 & &
    0.373 & 0.118 & & & \\
  NNPDF31\_nnlo\_pch\_as\_0116 & NNPDF NNLO 116 & &
    0.352 & 0.116 & & & \\
  NNPDF31\_nnlo\_pch\_as\_0120 & NNPDF NNLO 120 & &
    0.396 & 0.120 & & & \\
\hline
\end{tabular}
\end{center}
\caption{\label{tab:pdfs}  Overview of the PDF sets considered in the present study. It is understood that the number of active quark flavours is $n_f=3$ for $\as(1.3 \gev)$ and $n_f=5$ for $\as(M_Z)$.  References and quark-mass values are the same for PDF sets at different orders.  The same holds for $x_{\text{min}}$, which is the smallest momentum fraction for which PDF values are given by LHAPDF.}
\end{sidewaystable}

%%%%%%%%%%%%%%%%%%%%%%%%%%%%%%%%%%%%%%%%%%%%%%%%%

\subsection{PDF moments and \texorpdfstring{$x$}{x} ranges}
\label{sec:moments-x}

\begin{figure*}
\begin{center}
\subfigure[$G(1.5)$]{\includegraphics[height=12em,trim=20 0 170 0,clip]{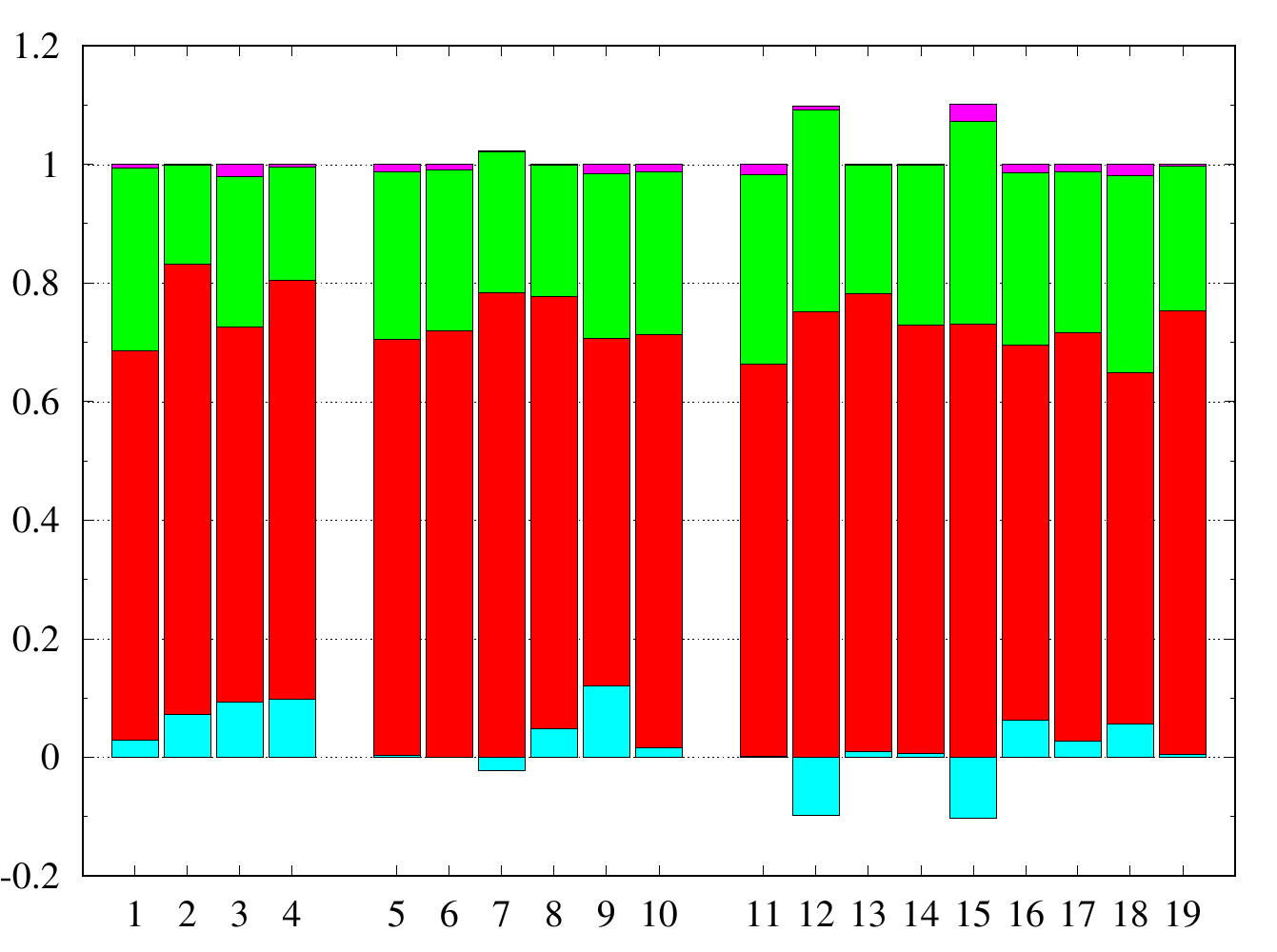}}
\subfigure[$\Qbar(1.5)$ \hspace{6em}]{\includegraphics[height=12em]{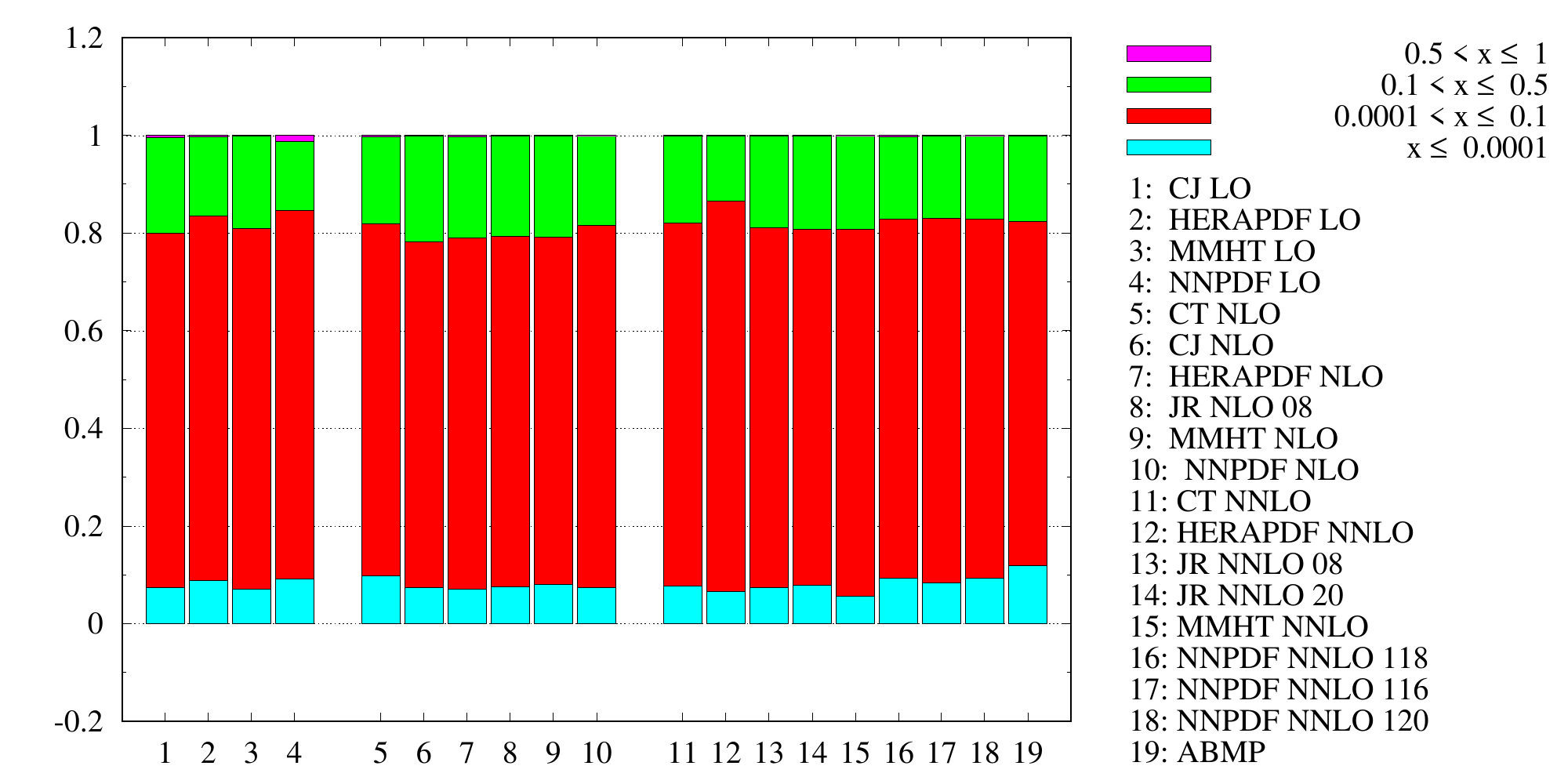}}
\\[2em]
\subfigure[$G(2)$]{\includegraphics[height=12em,trim=20 0 170 0,clip]{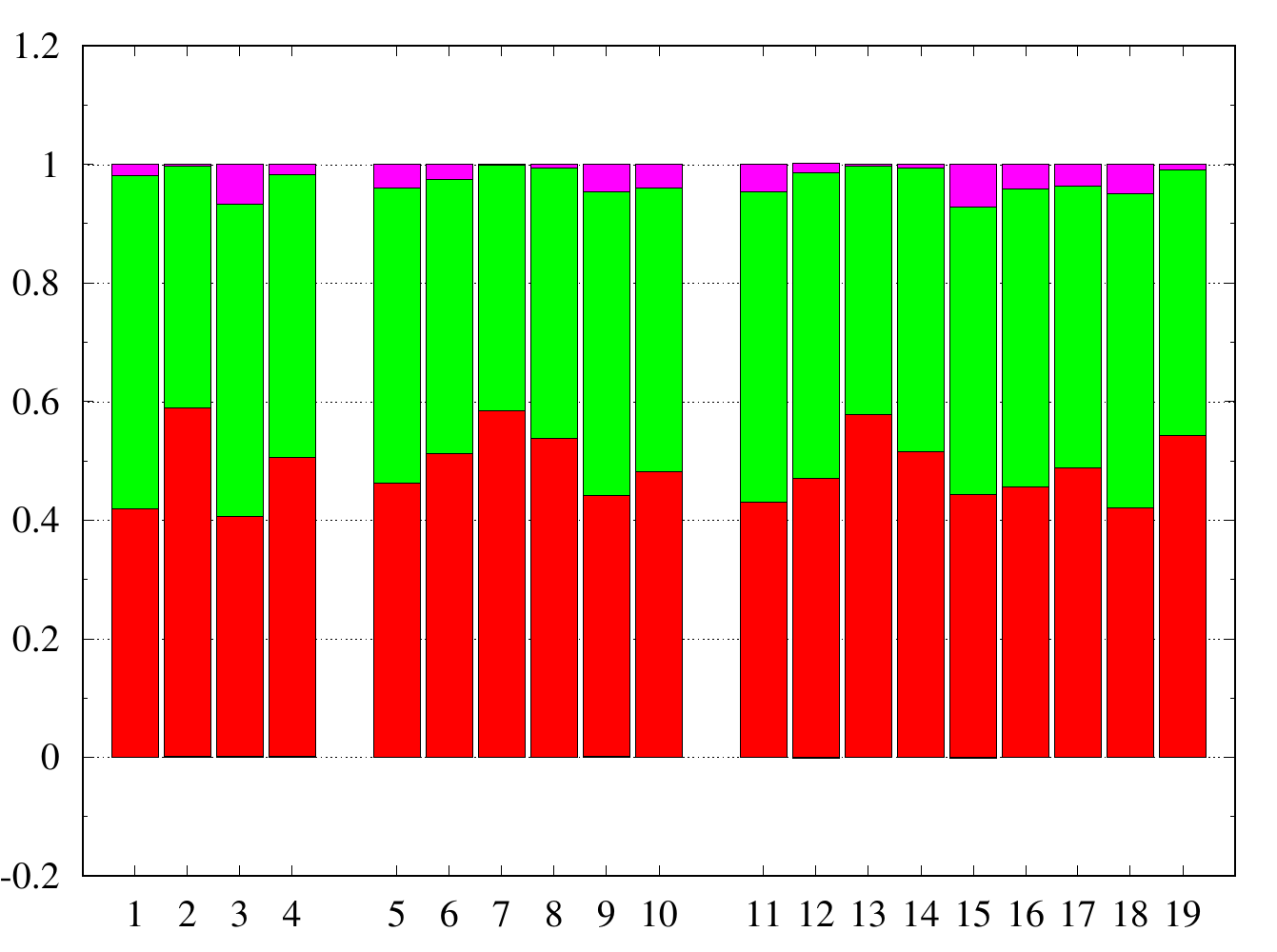}}
\subfigure[$\Qbar(2)$ \hspace{6em}]{\includegraphics[height=12em]{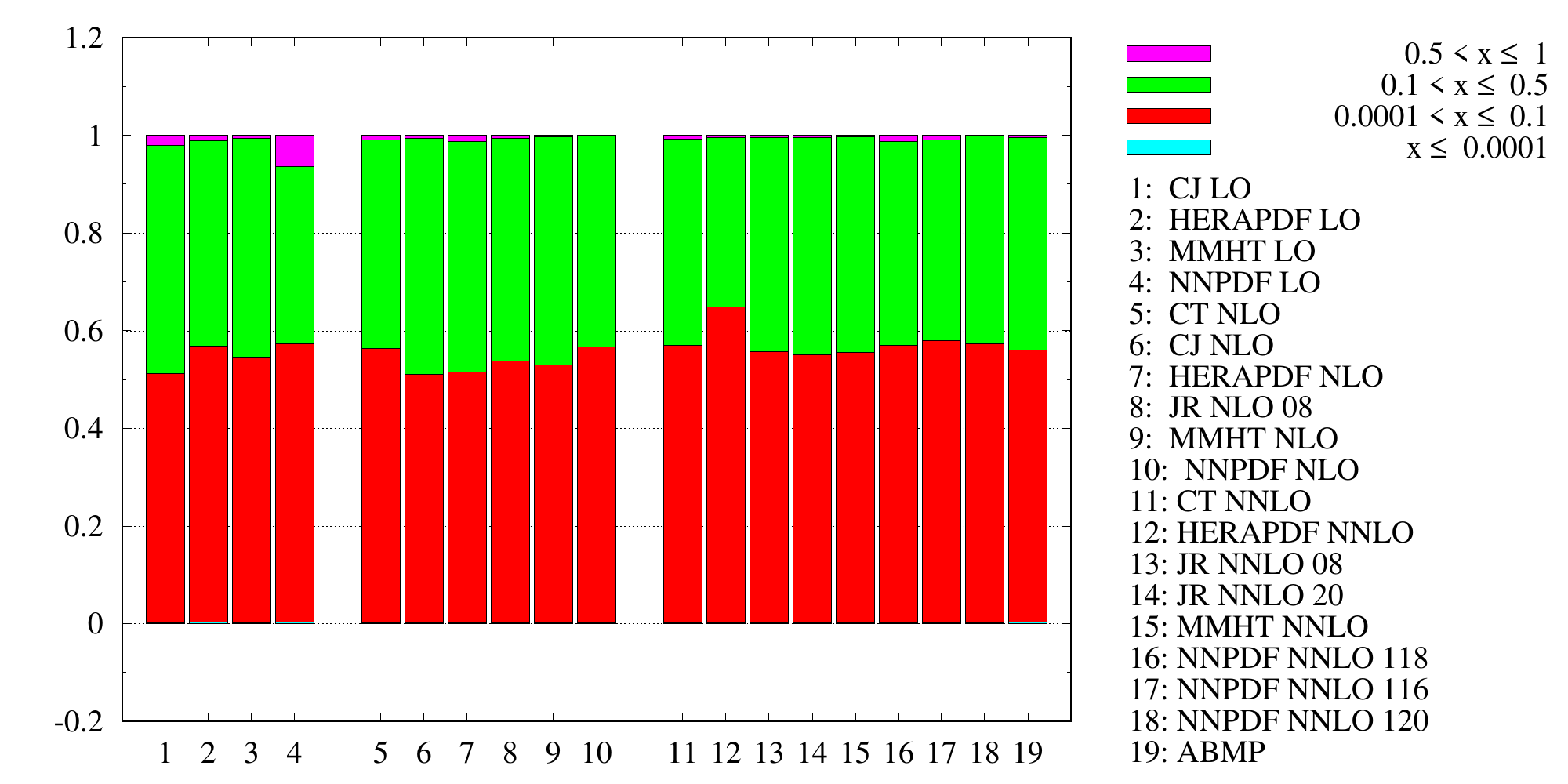}}
\\[1em]
\subfigure[$G(3)$]{\includegraphics[height=12em,trim=20 0 170 0,clip]{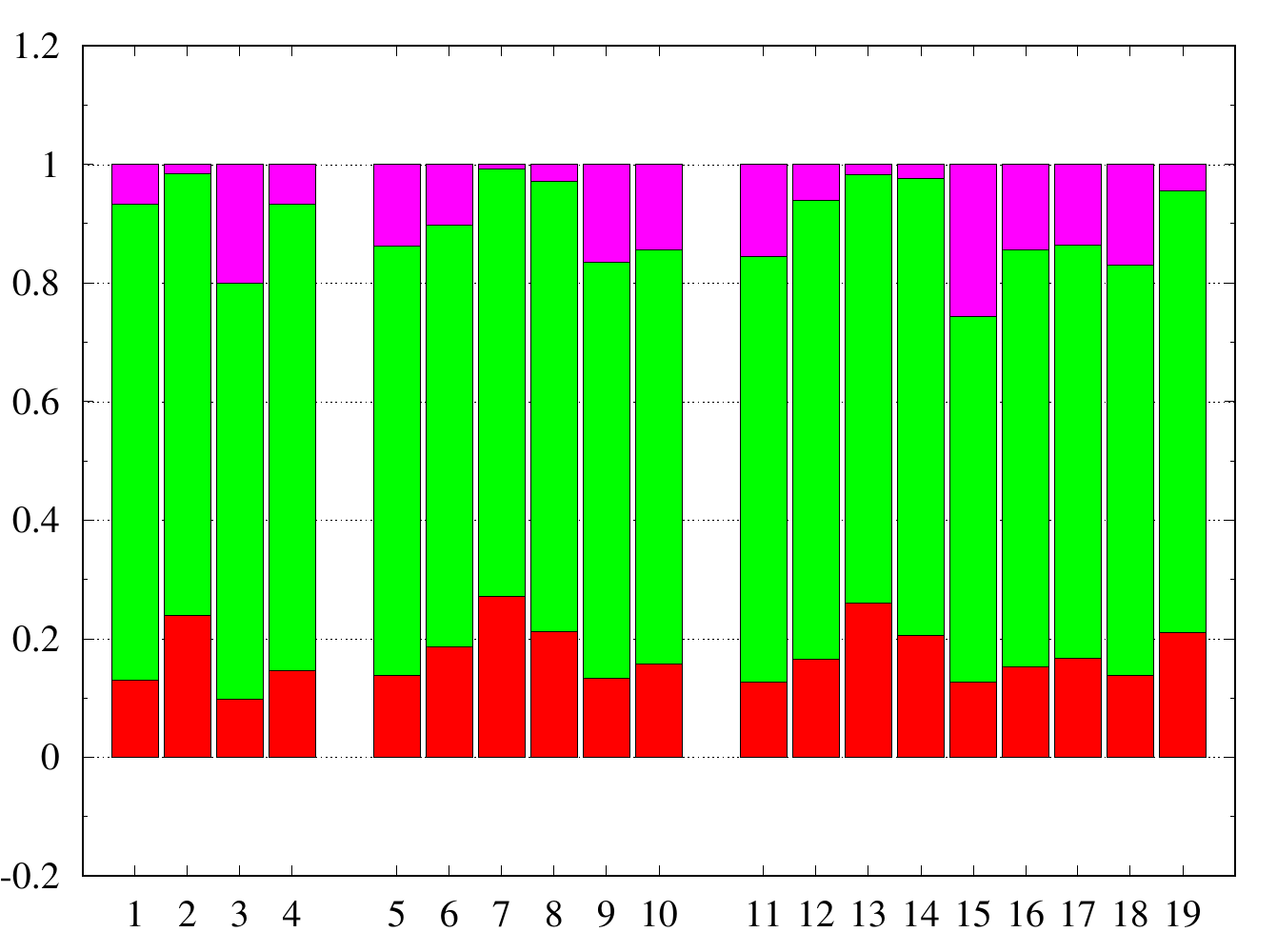}}
\subfigure[$\Qbar(3)$ \hspace{6em}]{\includegraphics[height=12em]{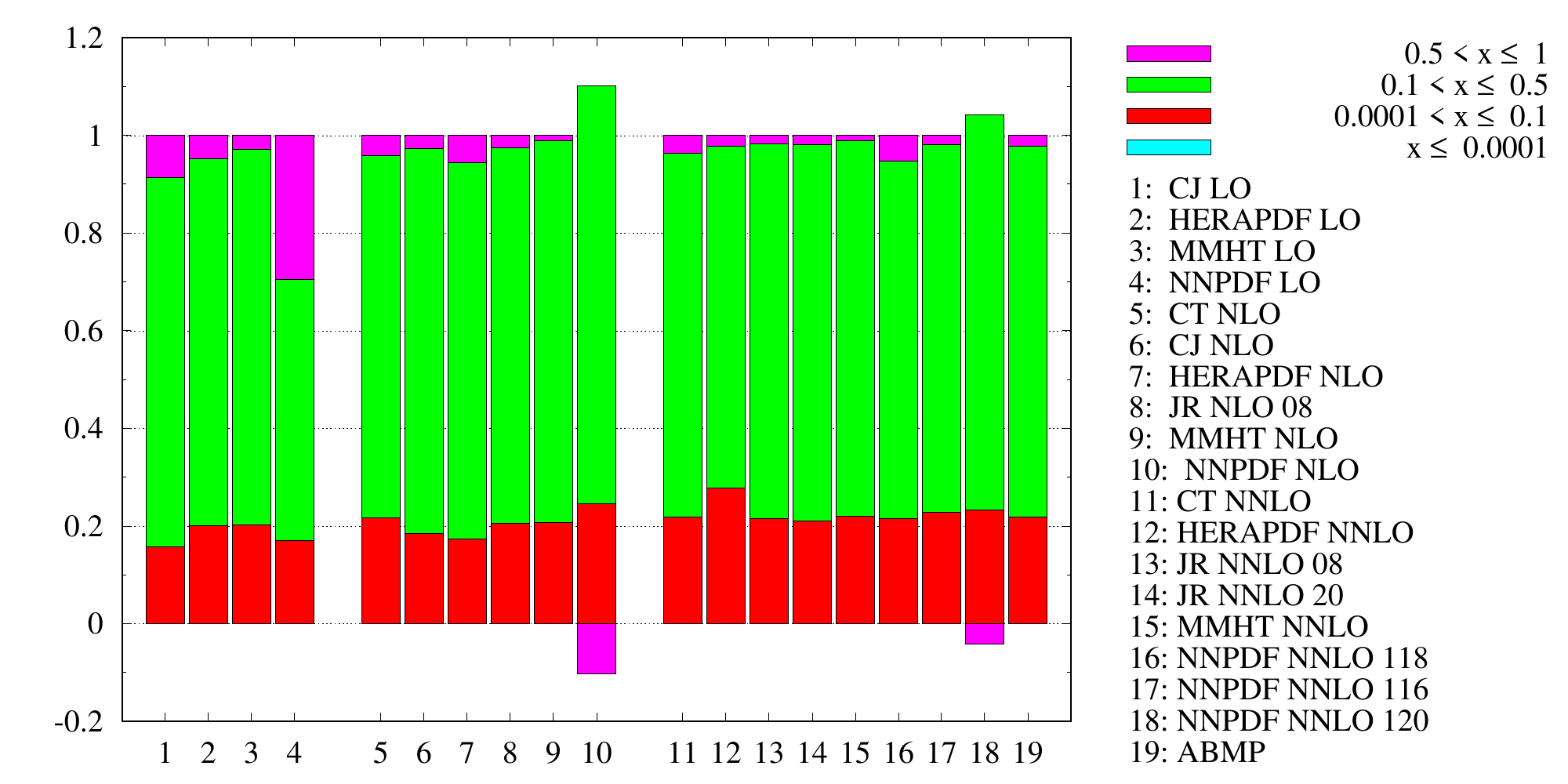}}
\end{center}
\caption{\label{fig:moments-x} Fractional contribution of different $x$ intervals to selected Mellin moments.  In some cases, the very small $x$ or the large $x$ region gives a negative contribution, indicating that the corresponding PDFs cannot be interpreted as number densities in that region.  The PDF sets are grouped according to their perturbative order (LO: 1--4, NLO: 5--10, NNLO: 11--19).}
\end{figure*}

We consider a collection of moments, from $j=1.5$ to $j=3$ in steps of $0.5$, in order to be sensitive to PDFs in a wide range of momentum fractions $x$.  To quantify this sensitivity, we divide the $x$ range into four intervals,
\begin{align}
x < 10^{-4} \,, \qquad 10^{-4} < x < 0.1 \,,
\qquad 0.1 < x < 0.5 \,, \qquad 0.5 < x \,,
\end{align}
to which we respectively refer as ``very small $x$ region'', ``small $x$ region'', ``valence region'', and ``large $x$ region'' in the sequel.  We then take the PDFs at the starting scale $\mu_0$ of our study and determine the contribution of these different $x$ intervals to each Mellin moment.  The result is shown for selected moments in \fig{\ref{fig:moments-x}} and summarised in \tab{\ref{tab:moments-x}}.

\begin{table}
\begin{center}
\begin{tabular}{|cl|cccc|} \hline
  moment & $j$ &
  $x < 10^{-4}$ & $10^{-4} < x < 0.1$ & $0.1 < x < 0.5$ & $x> 0.5$
  \rule[-0.6em]{0pt}{1.8em} \\ \hline
  $G$ \rule{0pt}{1.2em}
  & 1.5 & $-10\%$ to 10\% & 70\% to 80\% & $< 30\%$ & \\
  & 2   & & 40\% to 60\% & 40\% to 60\% & few \% \\
  & 2.5 & & 20 \% to 40\% & 60\% to 70\% & $< 15\%$ \\
  & 3   & & $< 15\%$ & 70\% to 80\% & $< 25\%$  \rule[-0.6em]{0pt}{1em}
  \\ \hline
  $\Qbar$ \rule{0pt}{1.2em}
  & 1.5 & $< 10\%$ & $\sim 70\%$ & $< 20\%$ & \\
  & 2   & & $\sim 60\%$ & $\sim 40\%$ & \\
  & 2.5 & & $\sim 40\%$ & $\sim 60\%$ & few \% \\
  & 3   & & $\sim 20\%$ & $\sim 80\%$ & few \%  \rule[-0.6em]{0pt}{1em}
  \\ \hline
  $Q$ \rule{0pt}{1.2em}
  & 1.5 & few \% & $\sim 40\%$ & $\sim 55\%$ & $\sim 5\%$ \\
  & 2   & & $\sim 20\%$ & $\sim 65\%$ & $\sim 15\%$ \\
  & 2.5 & & $\sim 10\%$ & $\sim 70\%$ & $\sim 20\%$ \\
  & 3   & & few \% & $\sim 65\%$ & $\sim 30\%$ \rule[-0.6em]{0pt}{1em}
  \\ \hline
\end{tabular}
\end{center}
\caption{\label{tab:moments-x}  Fractional contribution of different $x$ intervals to Mellin moments of PDFs.  Numbers are rounded to multiples of $5\%$ and are indicative for all sets considered in our study.  A blank entry indicates a negligible contribution.  Not included in this compilation are the following antiquark moments from NNPDF (which have relatively large errors): for the LO set, the region $x>0.5$ contributes about 20\% to $j=2.5$ and 30\% to $j=3$, whereas the NLO and NNLO sets give negative contributions from the same region to $j=3$ (about $-10\%$ for NLO and $-5\%$ for NNLO).}
\end{table}

Notice that the gluon moments have the strongest variation between different sets, with less variation for antiquarks and even less for quarks.  We see that $G(2)$ and $\Qbar(2)$ receive comparable contributions from the small $x$ region and the valence region.  To have quantities that are dominated by the small $x$ region, we resort to non-integer moments with $j=1.5$ (the next smallest integer $j=1$ is not an option, because the corresponding Mellin integrals are in general divergent for $G$, $\bar{Q}$, and $Q$).  We note that the very small $x$ region, in which fitted PDFs are essentially unconstrained by data, plays only a minor role in all moments we consider.  The high moments with $j=2.5$ and $j=3$ are increasingly dominated by the valence region.  The region of large $x$, where gluon and antiquark distributions are again poorly known, plays only a minor role in the $G$ and $\Qbar$ moments.  To summarise, we find that at our starting scale $\mu_0 = 1.3 \gev$, the Mellin moments we consider offer a reasonably differential sensitivity to the region $10^{-4} < x < 0.5$, in which PDFs are reasonably well known.  Moments with lower or higher $j$ will be more strongly affected by PDF uncertainties.  When going to lower scales, one should keep in mind that the PDFs are in general shifted towards higher $x$ values by backward evolution.  The relative importance of the different $x$ regions for a given $j$ will then change.

We note that for some PDF sets, the gluon distribution at $\mu_0 = 1.3 \gev$ becomes negative at very small $x$.  Some sets have a zero crossing of $g(x,\mu_0)$ at $x$ below $10^{-4}$.  For other sets (such as HERAPDF NNLO and MMHT NNLO) the zero crossing occurs already for $x$ below $10^{-3}$, and we see in \fig{\ref{fig:moments-x}} that the very small $x$ region gives a negative contribution to the lowest moment $G(1.5)$.  Such a behaviour is a clear indicator that the corresponding PDFs no longer admit a probability interpretation in the corresponding region.

When computing Mellin moments, we truncate the integral over $x$ at the smallest value $x_{\text{min}}$ for which LHAPDF provides PDF values for a given set, thus avoiding an extrapolation of parton densities down to $x=0$.  The values of $x_{\text{min}}$ are given in the last column of \tab{\ref{tab:pdfs}} and range from $10^{-9}$ to $10^{-6}$.  To obtain a conservative estimate of the uncertainty on the moments due to this truncation, we recompute them with a lower integration limit of $10 \, x_{\text{min}}$ instead of $x_{\text{min}}$.  The resulting change is less than $0.1 \%$ for all moments with $j \ge 2$.  For the $j = 1.5$ moments, it is less than $3 \%$ with the following exceptions.  The antiquark moment $\Qbar(1.5)$ of the NNPDF LO set changes by $4\%$, which is negligible compared with the huge error on this moment due to the PDF uncertainty at $x > x_{\text{min}}$.  The gluon moment $G(1.5)$ changes by $5\%$ for HERAPDF NNLO and by $4\%$, $8\%$, and $7\%$ for the MMHT sets at LO, NLO, and NNLO, respectively.  Our conclusions in \sect{\ref{sec:results}} are not affected by these somewhat larger truncation uncertainties.

When evolving the Mellin moments to lower scales, we use anomalous dimensions at fixed order in perturbation theory.  One may wonder whether some type of all-order resummation would improve perturbative convergence.  We argue that this is not the case: small-$x$ logarithms $\as \log(x)$ correspond to powers of $\as /j$ in Mellin space and hence do not require resummation for $j > 1$.  Large-$x$ logarithms, which correspond to powers of $\as \log^2 j$ \cite{Sterman:1986aj,Catani:1989ne}, do not appear in anomalous dimensions for the evolution of parton densities \cite{Korchemsky:1988si,Moch:2004pa,Vogt:2004mw}, and in any event, $\log j$ is not large for $j \le 3$.

%%%%%%%%%%%%%%%%%%%%%%%%%%%%%%%%%%%%%%%%%%%%%%%%%

\subsection{The running coupling}
\label{sec:alphas}

\begin{figure*}[ht]
\begin{center}
\subfigure[\label{fig:alphas-N4LO}]{\includegraphics[width=0.495\textwidth,trim=30 0 0 0,clip]{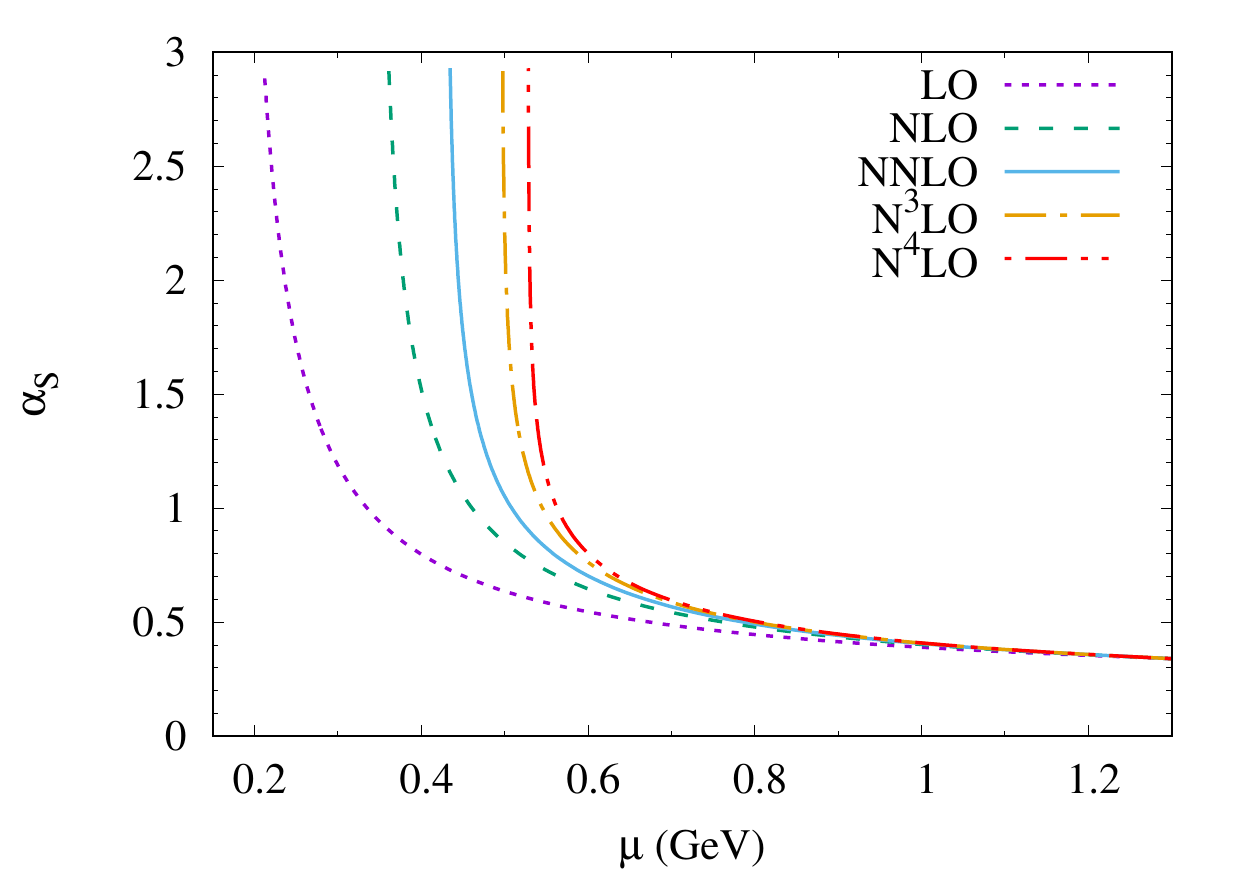}}
\hfill
\subfigure[\label{fig:alphas-NNLO}]{\includegraphics[width=0.495\textwidth,trim=30 0 0 0,clip]{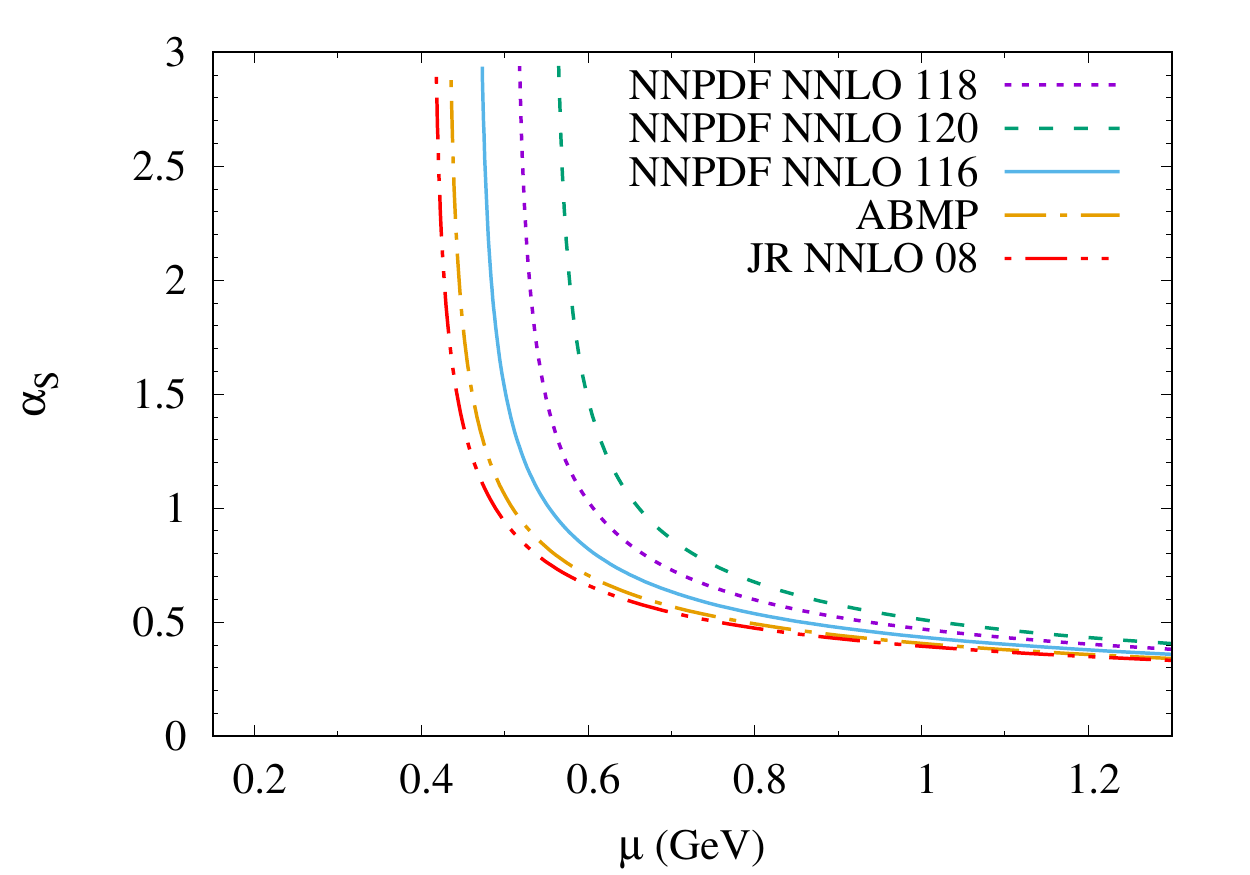}}
\end{center}
\caption{\label{fig:alphas} (a) The running coupling for $n_f = 3$ at different perturbative orders.  A common value $\as(1.3 \gev) = 0.378$ has been assumed.  (b) The running coupling for $n_f = 3$ corresponding to the values of $\as(1.3 \gev)$ taken in different NNLO parton sets.  As follows from \tab{\protect\ref{tab:pdfs}}, the curves for the remaining NNLO sets of our study are in-between those shown in the plot.}
\end{figure*}

Let us take a brief look at the running coupling $\as(\mu)$.  In \fig{\ref{fig:alphas-N4LO}} we show the evolution of $\as(\mu)$ down to low scales at different orders in the perturbative expansion of $\beta(\as)$, which is available up to five-loop order \cite{Baikov:2016tgj}, i.e.\ up to N$^4$LO.  For definiteness, we take in this plot a common value $\as(1.3 \gev) = 0.378$ for all orders (this is the value of the CT NNLO set as seen in \tab{\ref{tab:pdfs}}).  The plot looks very similar if instead one takes a common value $\as(M_Z) = 0.118$ for all orders and sequentially evolves and matches the coupling from $n_f=5$ down to $n_f=3$.  We observe that down to about $\mu \sim 0.7 \gev$, the difference between different orders is small and decreases with the order, indicating satisfactory convergence of the perturbative expansion.  For lower scales, however, the different orders differ more and more strongly.

\vspace{1em}

In \fig{\ref{fig:alphas-NNLO}} we show the running of $\as(\mu)$ at NNLO with starting values at $\mu_0 = 1.3\gev$ corresponding to those in different PDF sets.  We see that the moderate spread in $\as$ values at $\mu_0$ rapidly increases when evolving to lower scales.

\section{Evolution to low scales}
\label{sec:results}

We have now everything in place to investigate the evolution of Mellin moments for different $j$ from their starting values at $\mu_0 = 1.3\gev$ down to small scales.  We focus on the moments $G(j)$ and $\Qbar(j)$.  As a function of $\mu$, PDF moments exhibit a very steep behaviour at scales where $\as(\mu)$ starts to diverge, as is seen in \fig{\ref{fig:moments-sets}} below.  To display the low-scale behaviour of the moments in a clearer way, we will in general plot them as functions of $\as$ instead of $\mu$.

We remark that the lowest moments $G(1.5)$ and $\Qbar(1.5)$ are consistent with zero within huge errors for three sets, namely for NNPDF LO, CT NLO, and CT NNLO.  This already holds at $\mu_0$ and reflects the huge uncertainties of the antiquark and gluon distributions at low $x$ in these sets.  Evolving $G(1.5)$ and $\Qbar(1.5)$  to yet lower scales then gives no further useful information.  For simplicity, we will not explicitly mention these cases in the following discussion.

%%%%%%%%%%%%%%%%%%%%%%%%%%%%%%%%%%%%%%%%%%%%%%%%%%%

\subsection{Comparison of different orders}
\label{sec:compare-orders}

\begin{figure*}
\begin{center}
\subfigure[$\as \ms \gamma_{q q}(1.5) /\beta$]{\includegraphics[width=0.49\textwidth,trim=25 0 0 0,clip]{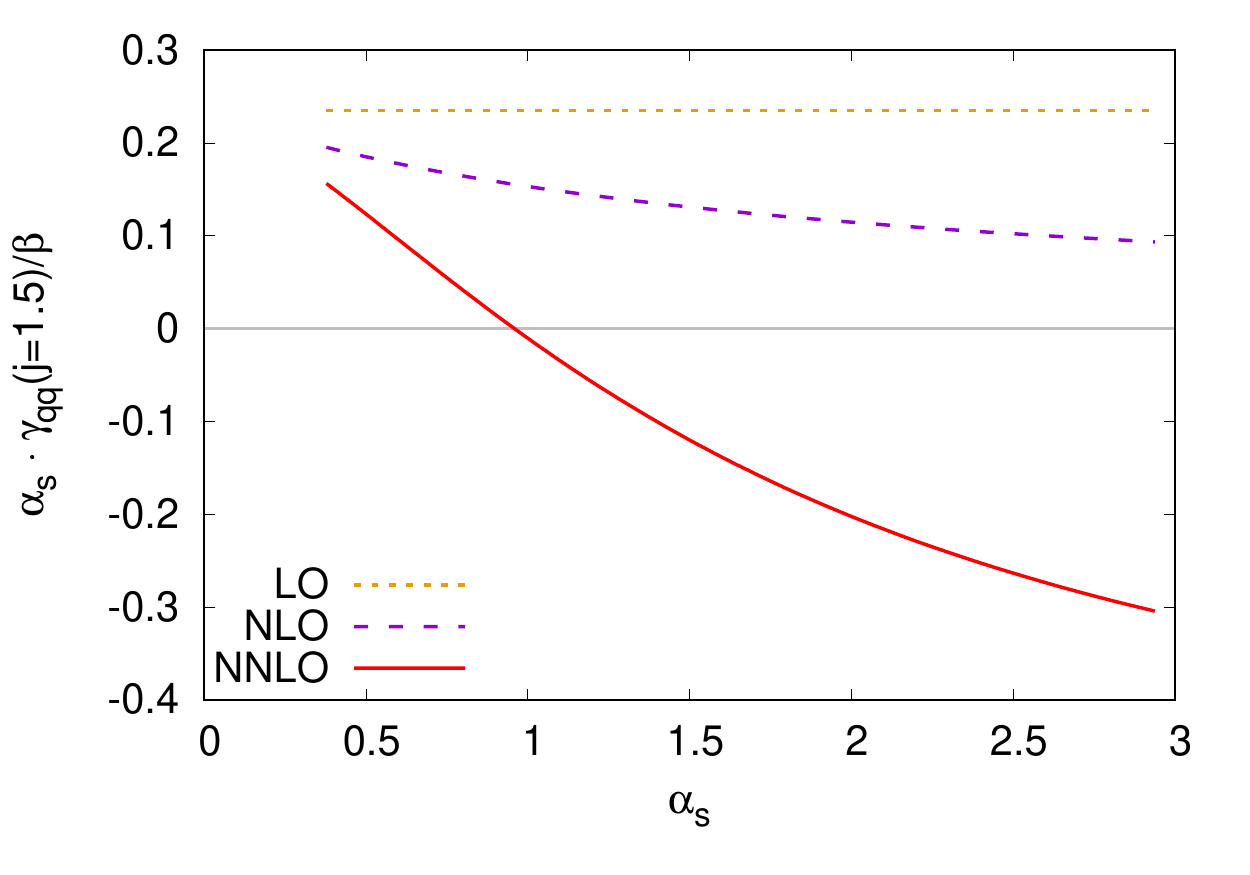}}
\hfill
\subfigure[$\as \ms \gamma_{\text{ns}}(1.5) /\beta$]{\includegraphics[width=0.49\textwidth,trim=25 0 0 0,clip]{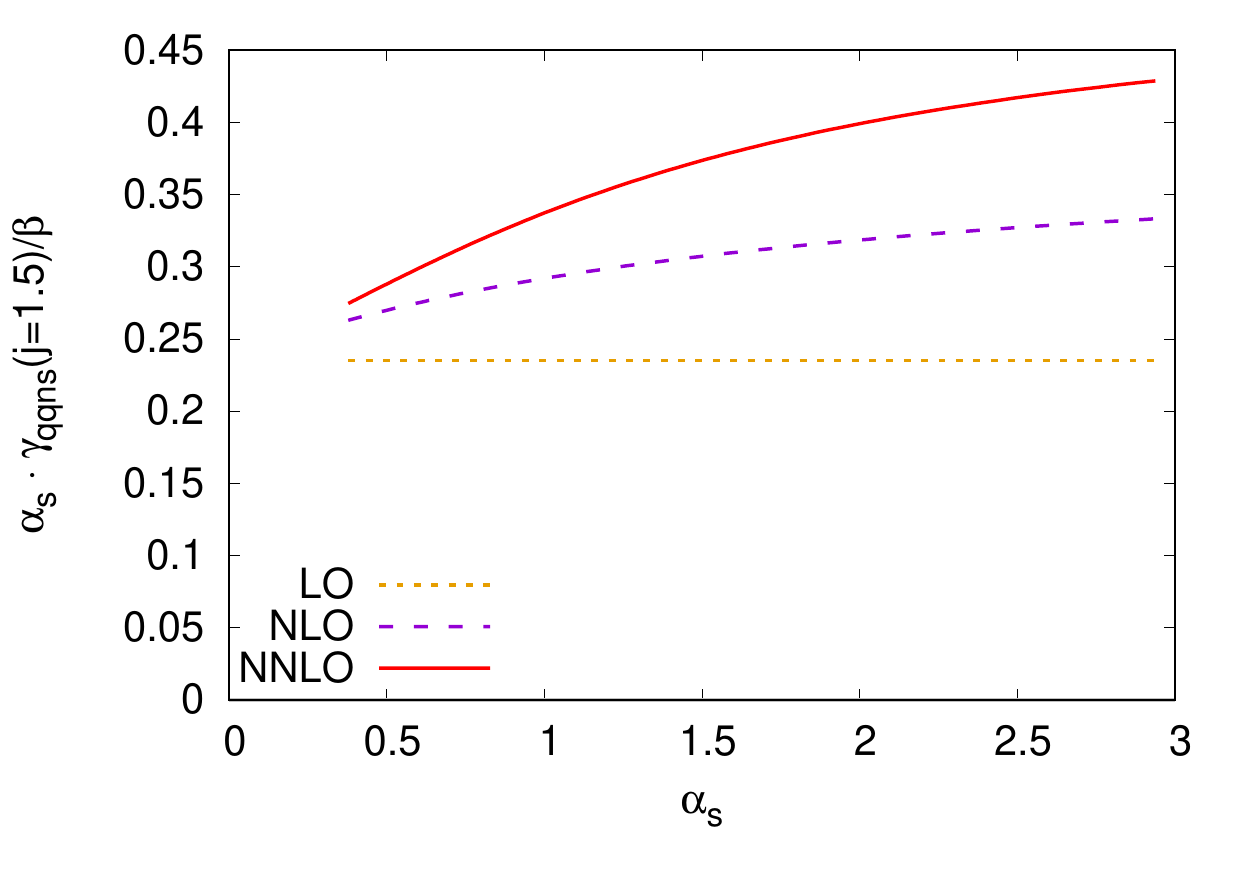}}
\\[1em]
\subfigure[$\as \ms \gamma_{q q}(2) /\beta$]{\includegraphics[width=0.49\textwidth,trim=25 0 0 0,clip]{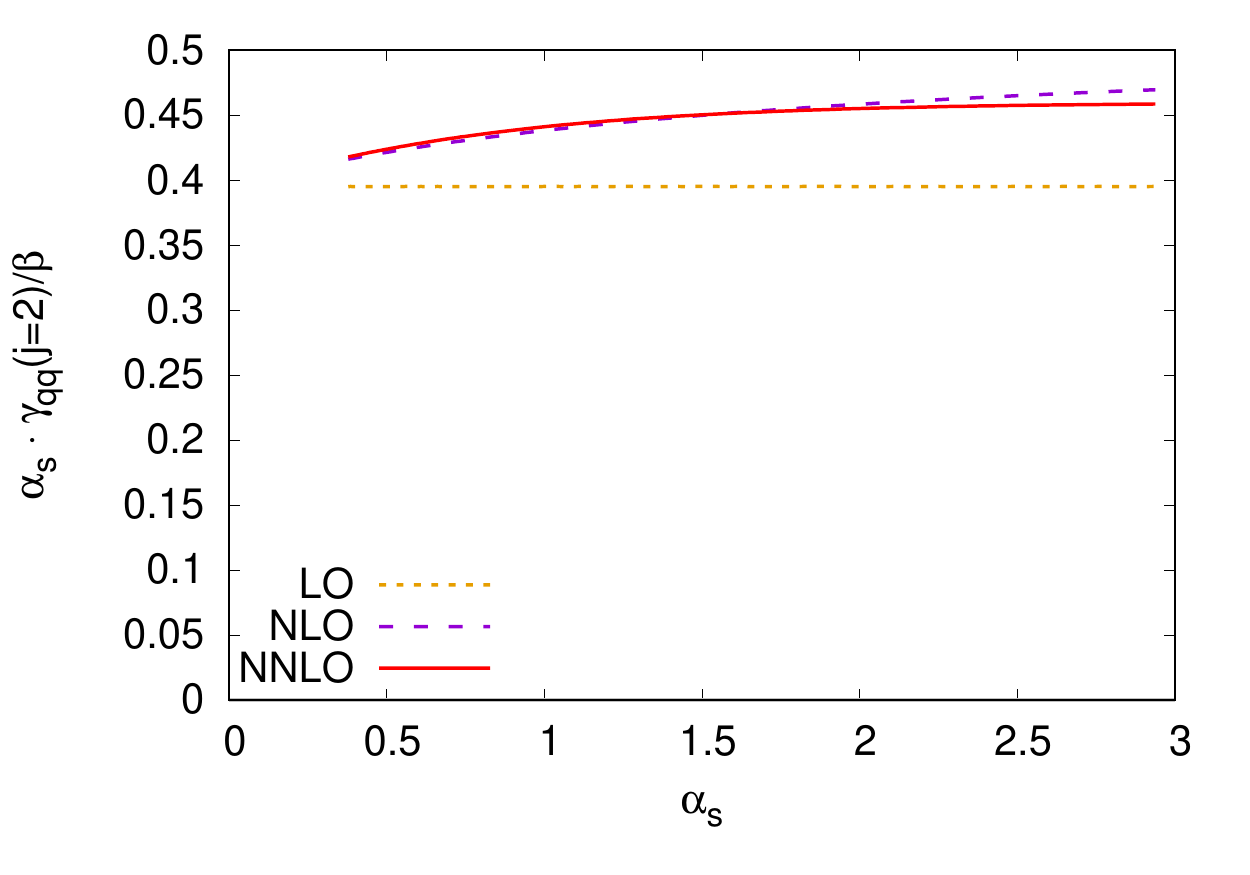}}
\hfill
\subfigure[$\as \ms \gamma_{q g}(3) /\beta$]{\includegraphics[width=0.49\textwidth,trim=25 0 0 0,clip]{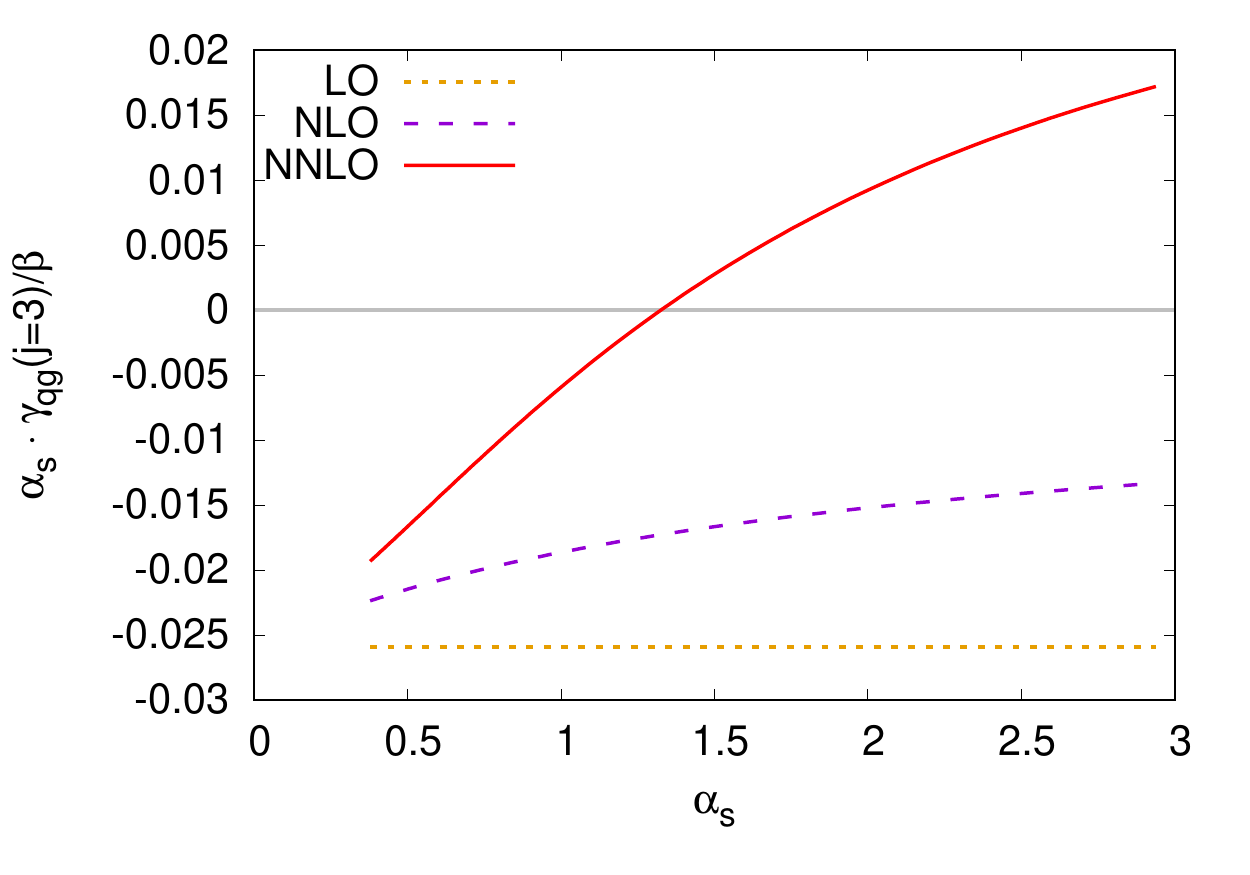}}
\\[1em]
\subfigure[$\as \ms \gamma_{g q}(2.5) /\beta$]{\includegraphics[width=0.49\textwidth,trim=25 0 0 0,clip]{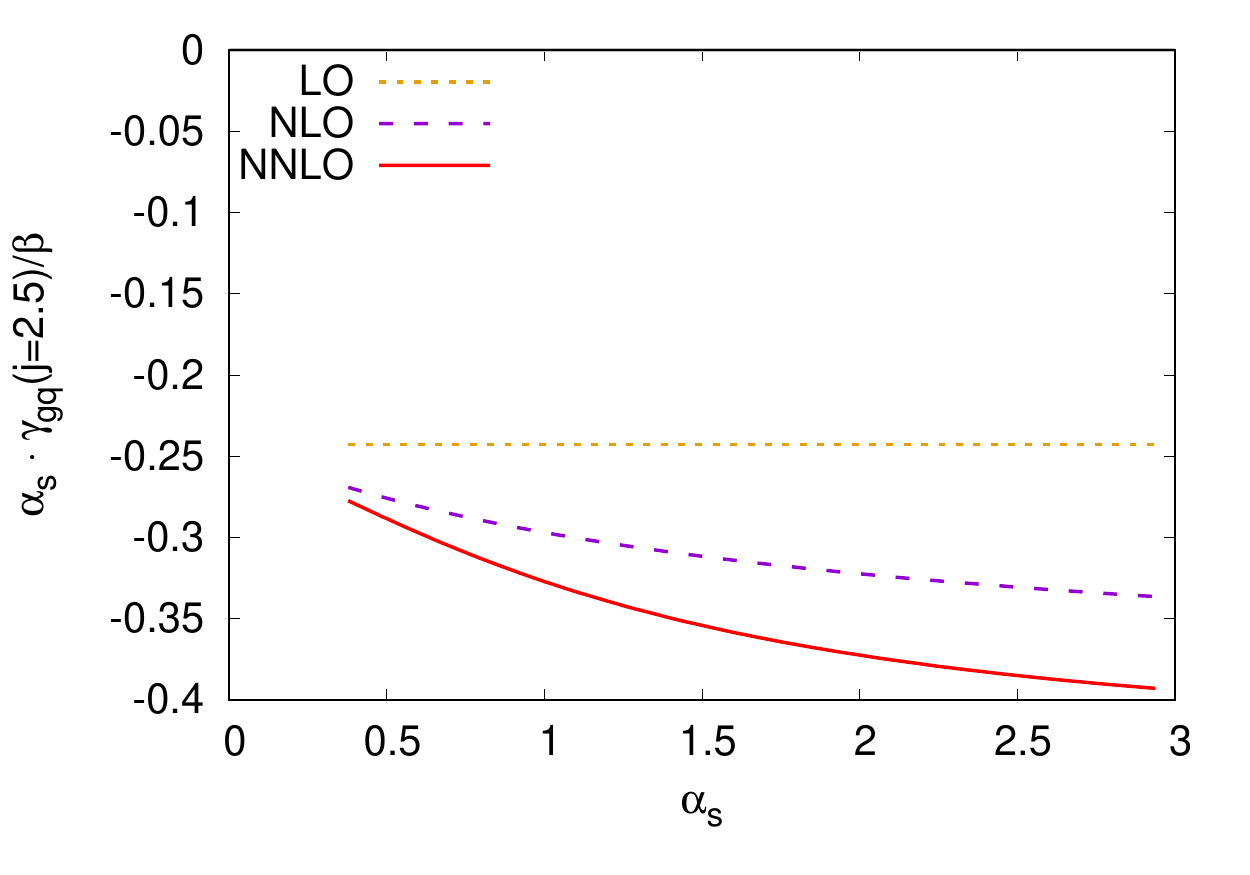}}
\hfill
\subfigure[$\as \ms \gamma_{g g}(2.5) /\beta$]{\includegraphics[width=0.49\textwidth,trim=25 0 0 0,clip]{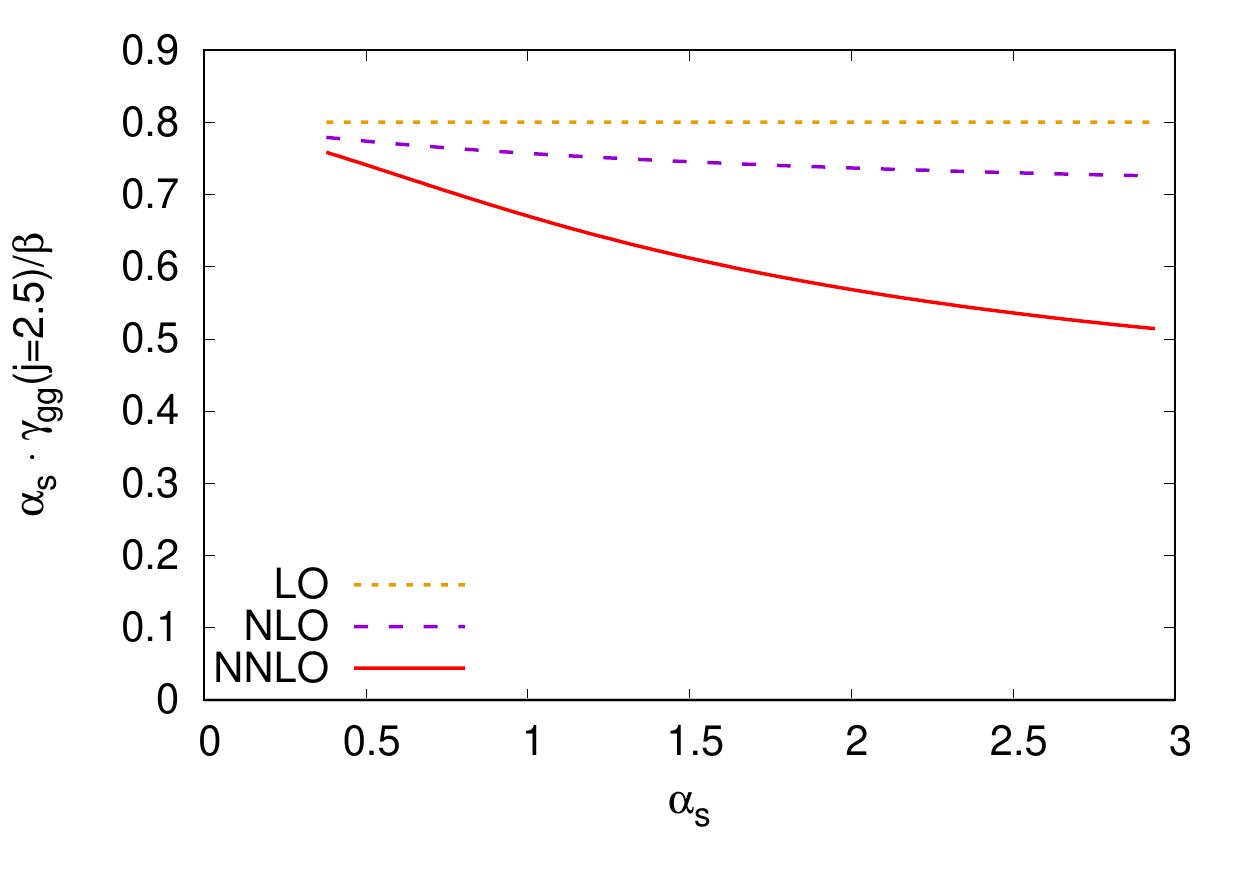}}
\end{center}
\caption{\label{fig:alpha-beta-gamma} Plots of $\as \ms \gamma_{i}(j,\as) / \beta(\as)$ at different perturbative orders for selected anomalous dimensions.
}
\end{figure*}

If we study PDF moments as functions of $\as$, then the quantity that drives their evolution is $\gamma_i(\as) /\beta(\as)$ according to \eqn{\eqref{nons-evo-alp}} and its analogue for the singlet channel.  In \fig{\ref{fig:alpha-beta-gamma}} we show this quantity for the three perturbative orders considered in our study, multiplied with an additional power of $\as$ so that the LO curves are constants.  The channels and moment indices $j$ shown in the figure are representative of the wide range of patterns we observe.  In some cases, there are only moderate differences between different orders, in other cases the difference between LO and NNLO amounts to a factor around two at the highest $\as$ shown in the figure, and in yet other cases one finds that the NNLO curve changes sign at some intermediate value of $\as$.  Let us emphasise that by showing the curves up to $\as = 3$, we do \emph{not} mean to imply that perturbation theory is valid up to that value.  Indeed, both figures \ref{fig:alphas} and \ref{fig:alpha-beta-gamma} suggest that for the quantities we are studying here, perturbation theory breaks down well before that value of $\as$ is reached.

\begin{figure*}
\begin{center}
\subfigure[$G(1.5)$]{\includegraphics[width=0.49\textwidth,trim=25 0 0 0,clip]{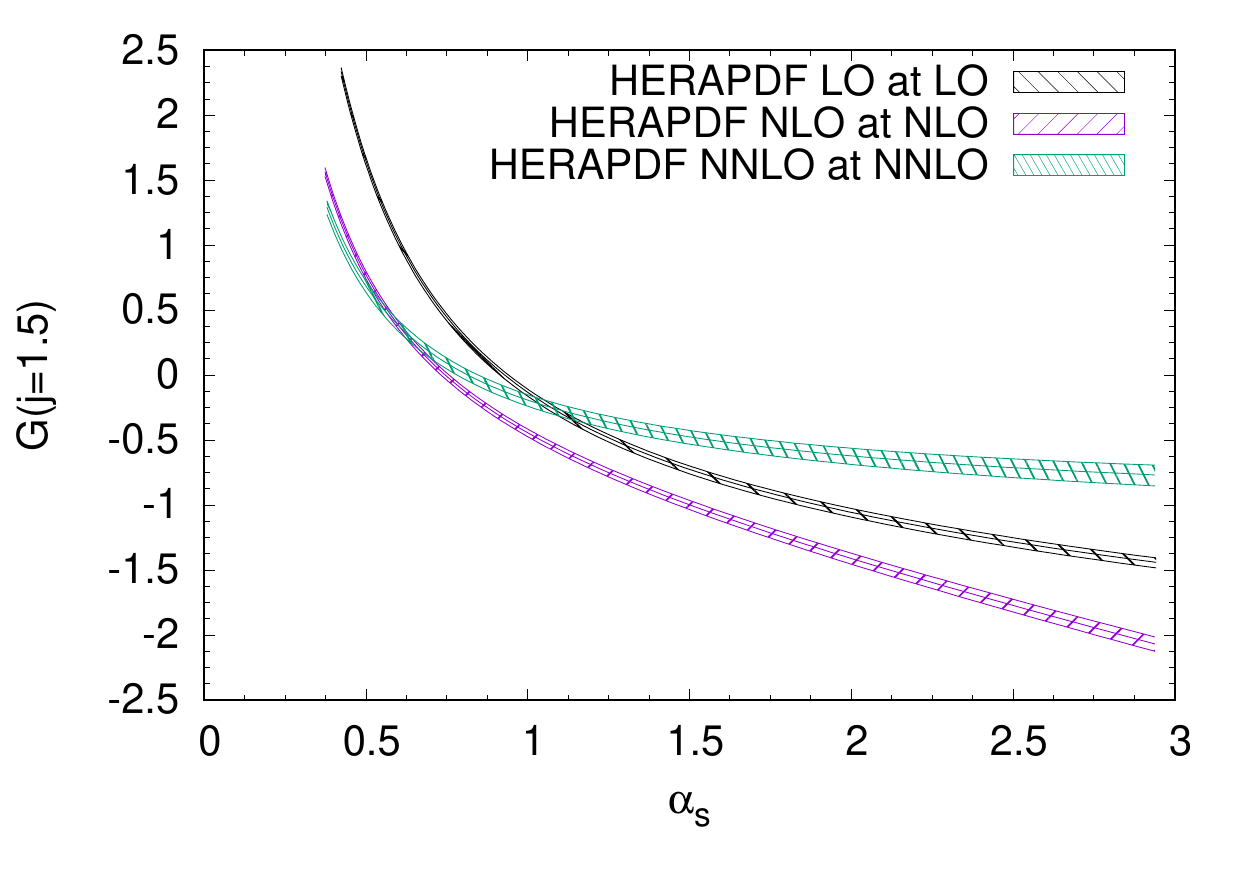}}
\hfill
\subfigure[$\Qbar(1.5)$]{\includegraphics[width=0.49\textwidth,trim=25 0 0 0,clip]{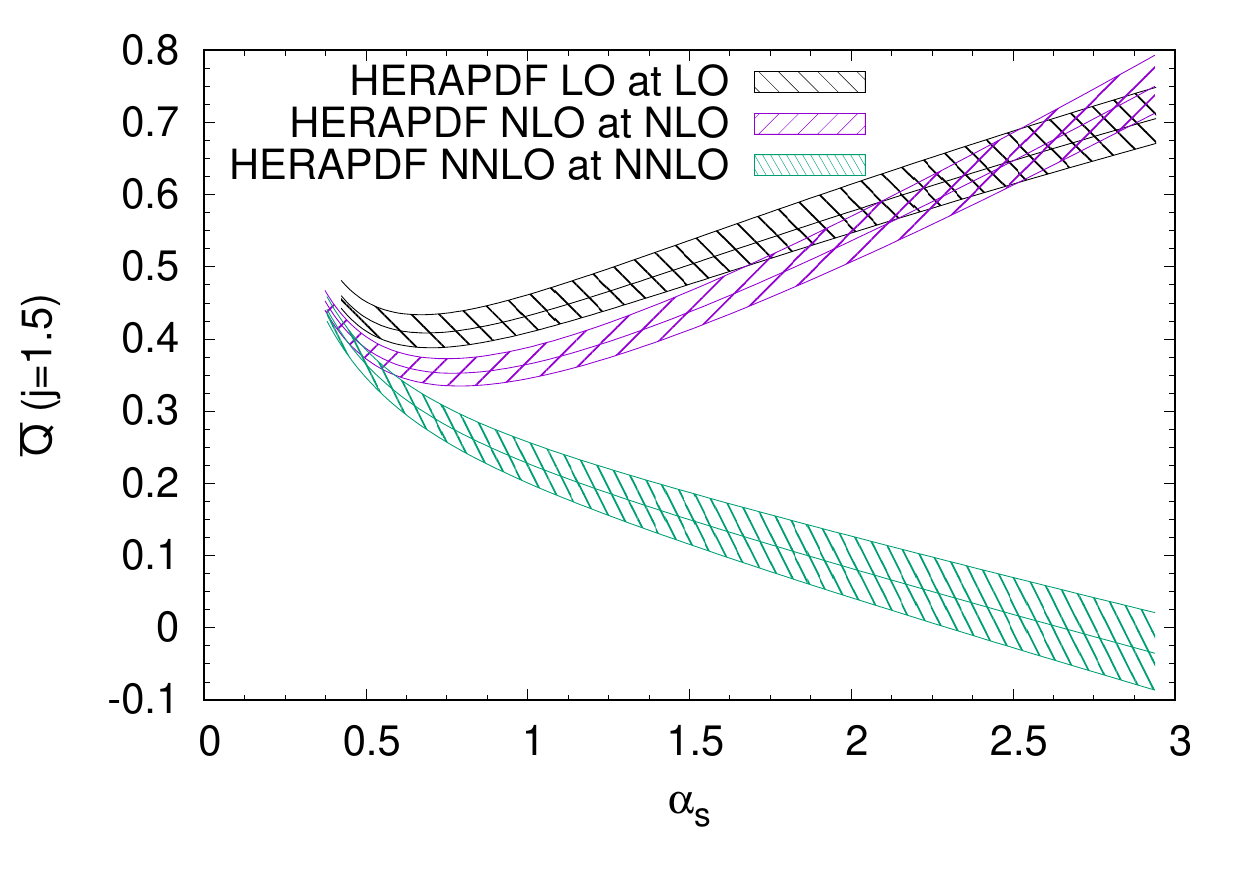}}
\\[1em]
\subfigure[$G(2.5)$]{\includegraphics[width=0.49\textwidth,trim=25 0 0 0,clip]{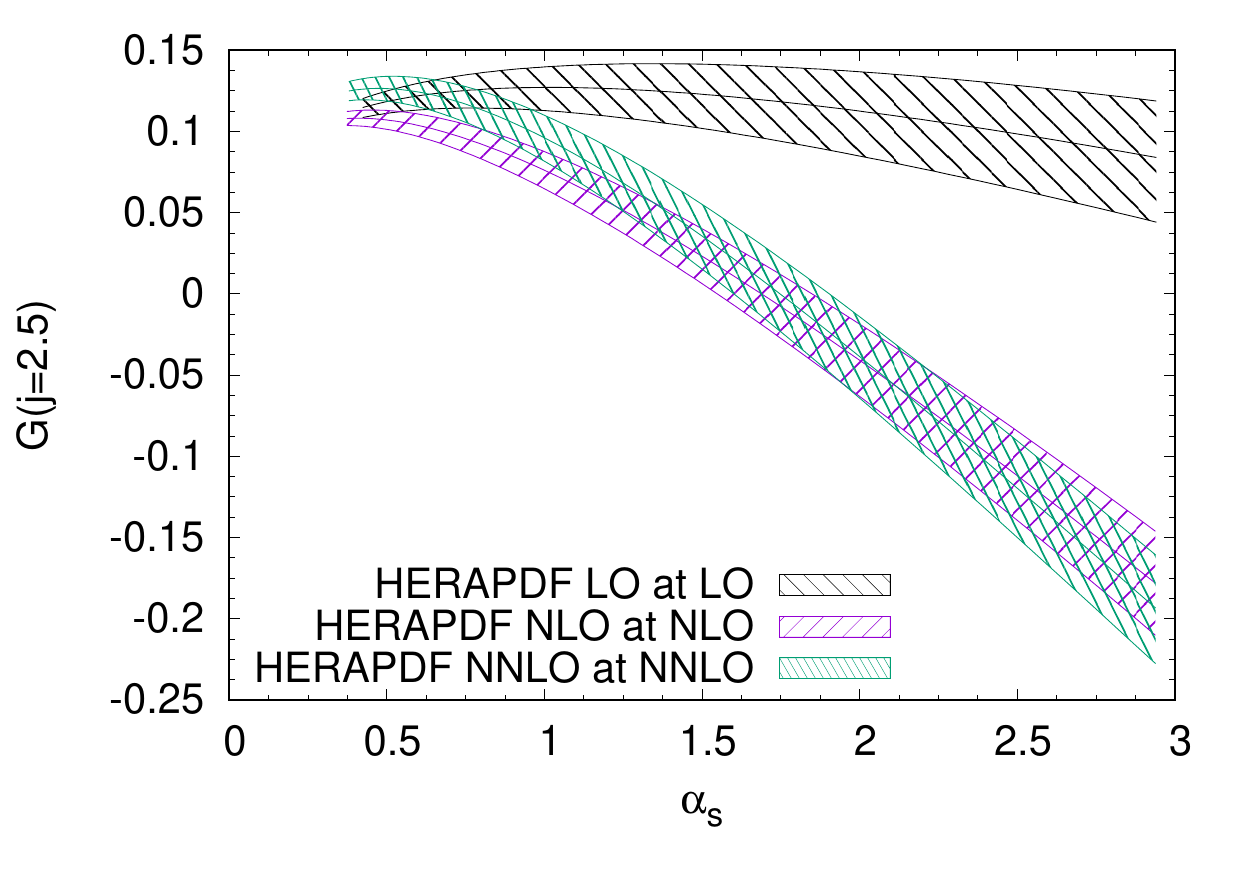}}
\hfill
\subfigure[$\Qbar(2.5)$]{\includegraphics[width=0.49\textwidth,trim=25 0 0 0,clip]{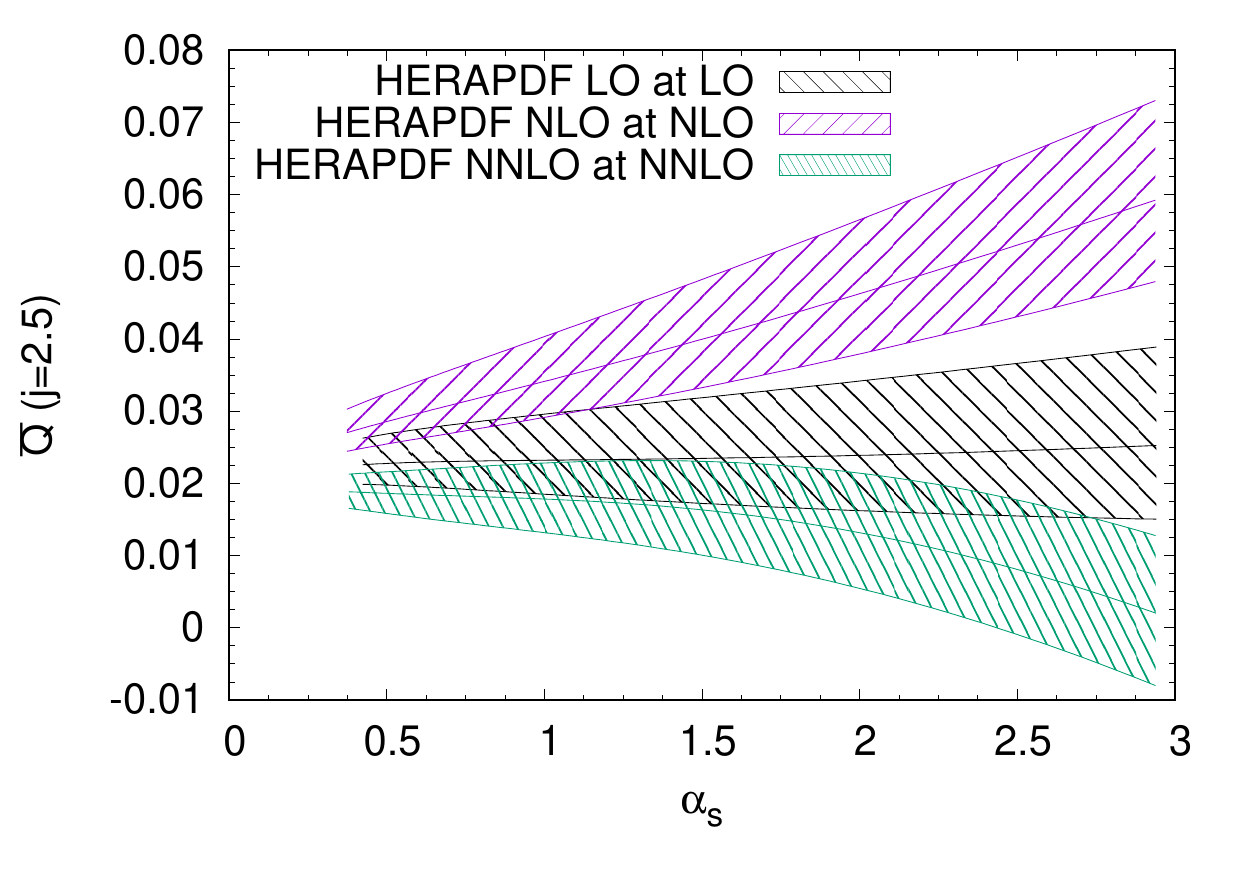}}
\\[1em]
\subfigure[$G(3)$]{\includegraphics[width=0.49\textwidth,trim=25 0 0 0,clip]{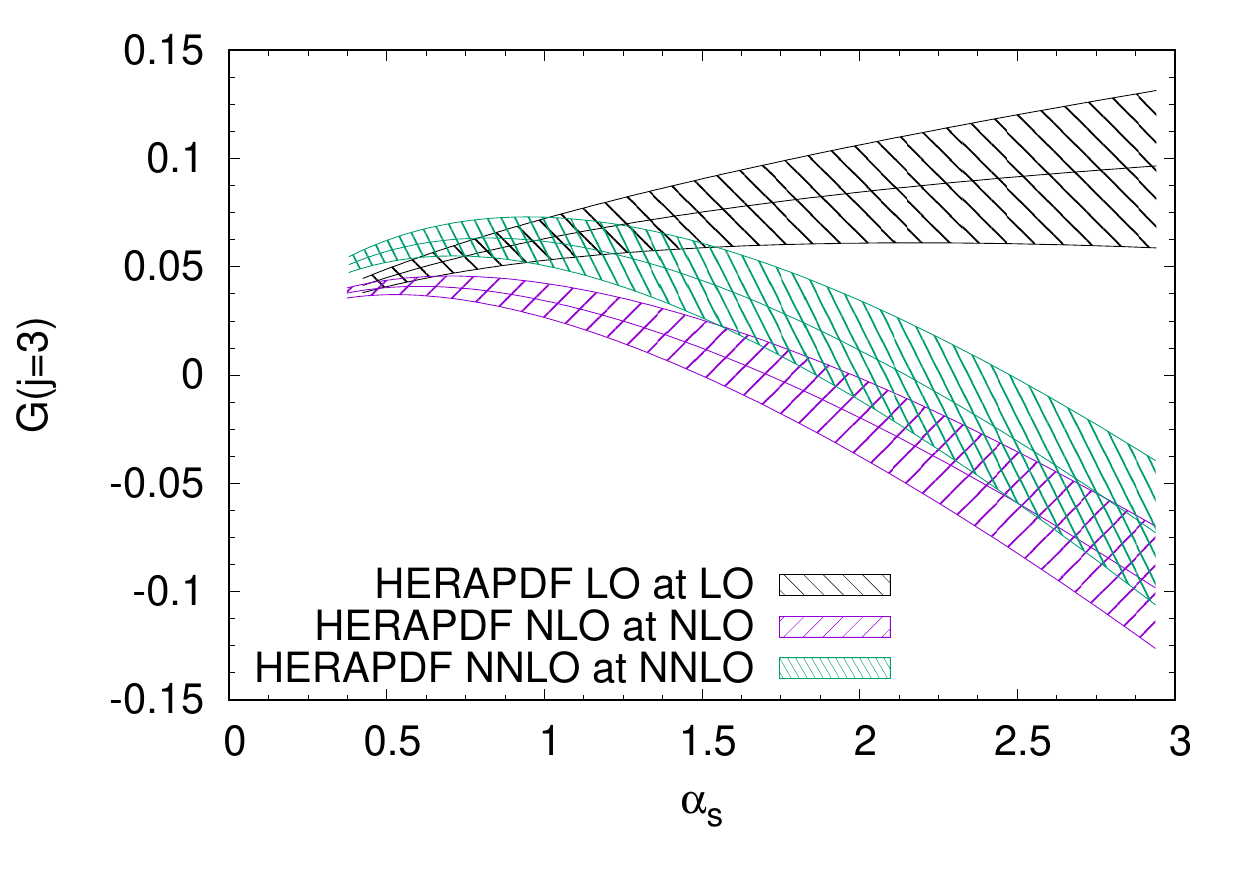}}
\hfill
\subfigure[$\Qbar(3)$]{\includegraphics[width=0.49\textwidth,trim=25 0 0 0,clip]{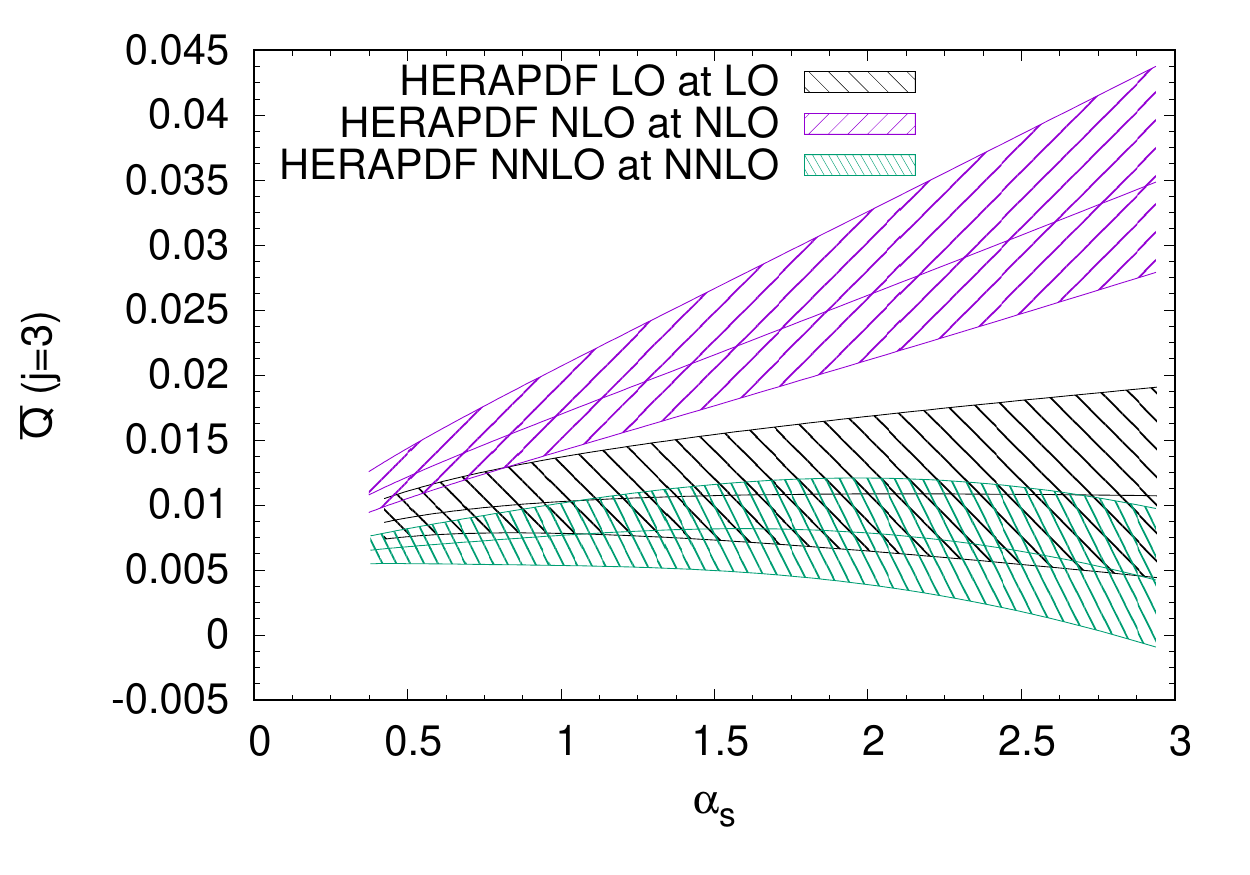}}
\end{center}
\caption{\label{fig:moments-orders} Comparison of moments computed for the HERAPDF sets at different perturbative orders.
}
\end{figure*}

To assess the difference in the evolution behaviour of Mellin moments at different orders, we focus on those sets that provide PDFs at all three orders, i.e.\ on HERAPDF, MMHT and NNPDF (see \tab{\ref{tab:pdfs}}).  A selection of moments for the HERAPDF sets is shown in \fig{\ref{fig:moments-orders}}.  For $G(j)$ with $j \ge 2$, we find the NLO and NNLO moments to be rather close to each other, while the LO moments are clearly different from these at large $\as$.  For $G(1.5)$, all orders differ quite noticeably.  The situation is similar for MMHT and NNPDF.  For the antiquark moments, we find in general larger differences between different orders, with a pattern that depends on $j$ and also on the considered PDF set.  This stronger dependence on the initial conditions is perhaps not surprising, given that $\Qbar$ is the difference between the combinations $(Q + \Qbar) /2$ and $(Q - \Qbar) /2$, which evolve independently of each other.  We note that the width of the error bands for the moments increases with $\as$, but in most cases it does so rather slowly.  The same is observed for the other PDF sets and is consistent with the numerical stability of backward evolution for Mellin moments.

%%%%%%%%%%%%%%%%%%%%%%%%%%%%%%%%%%%%%%%%%

\subsection{Comparison of different PDF sets}
\label{sec:compare-sets}

\begin{figure*}
\begin{center}
\subfigure[$G(2)$ at LO]{\includegraphics[width=0.49\textwidth,trim=25 0 0 0,clip]{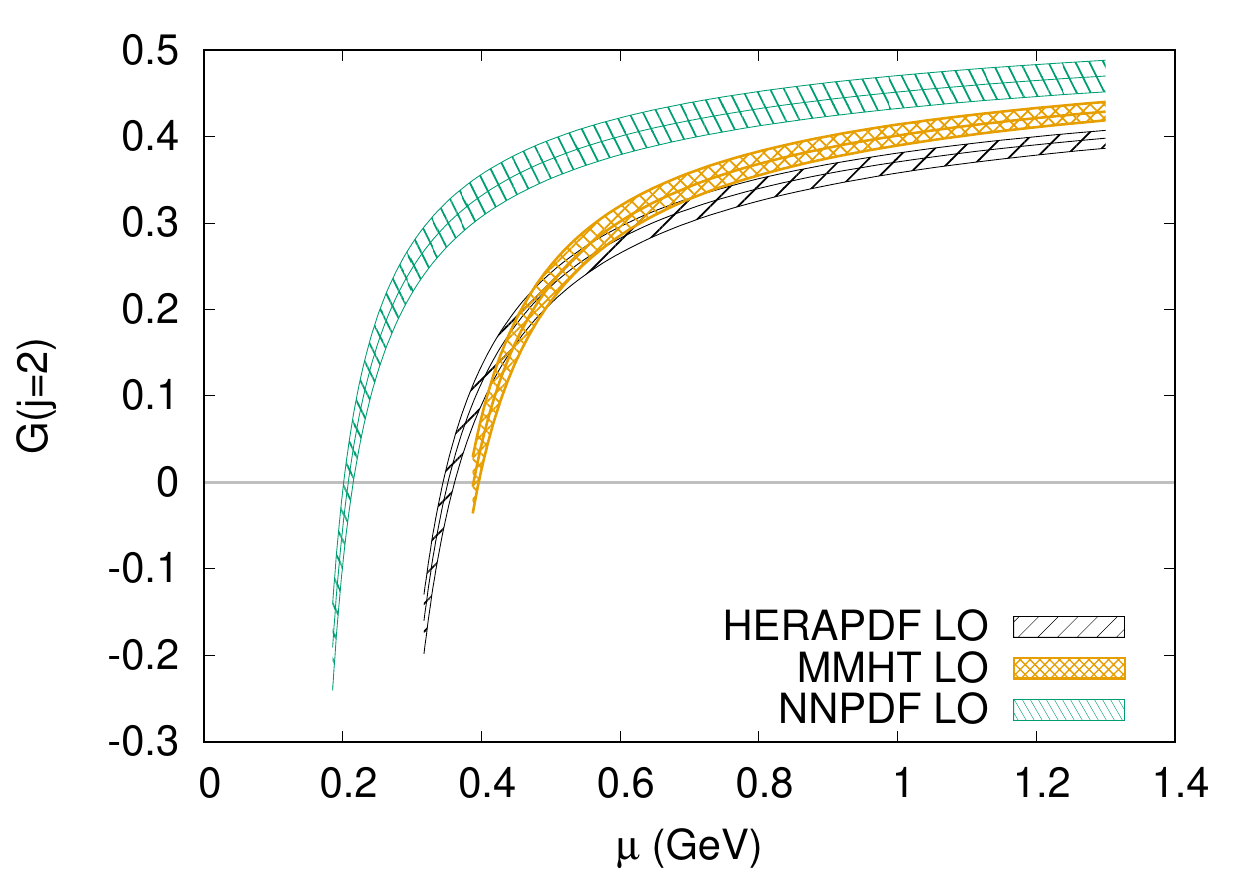}}
\hfill
\subfigure[\label{fig:mom-qbar-LO} $\Qbar(2)$ at LO]{\includegraphics[width=0.49\textwidth,trim=25 0 0 0,clip]{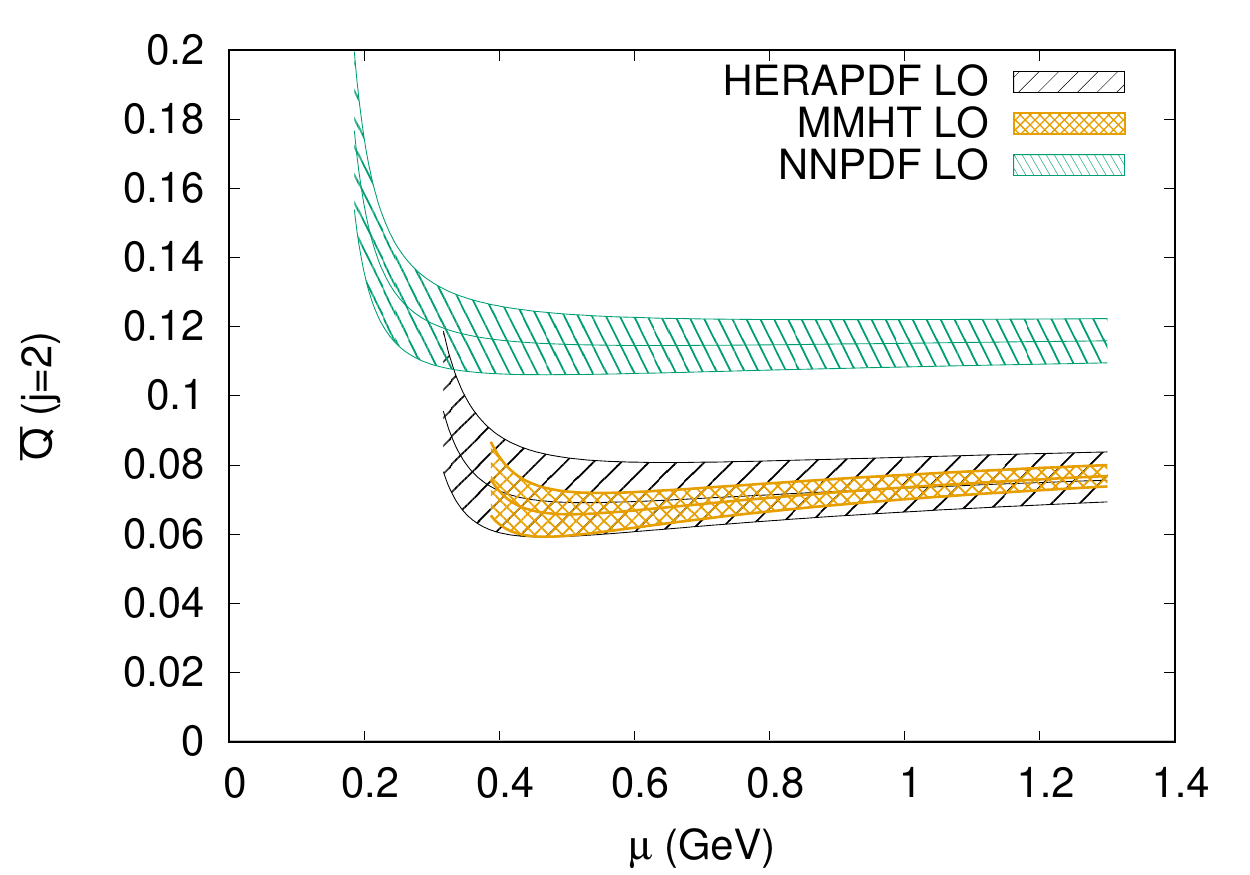}}
\end{center}
\caption{\label{fig:moments-sets-LO} Comparison of $j=2$ moments for gluons and antiquarks in the LO sets. The lowest value of $\mu$ shown corresponds to  $\alpha_s(\mu) = 3$.
}
\end{figure*}

\begin{figure*}
\begin{center}
\subfigure[\label{fig:mom-glu-LO} $G(3)$ at LO]{\includegraphics[width=0.49\textwidth,trim=25 0 0 0,clip]{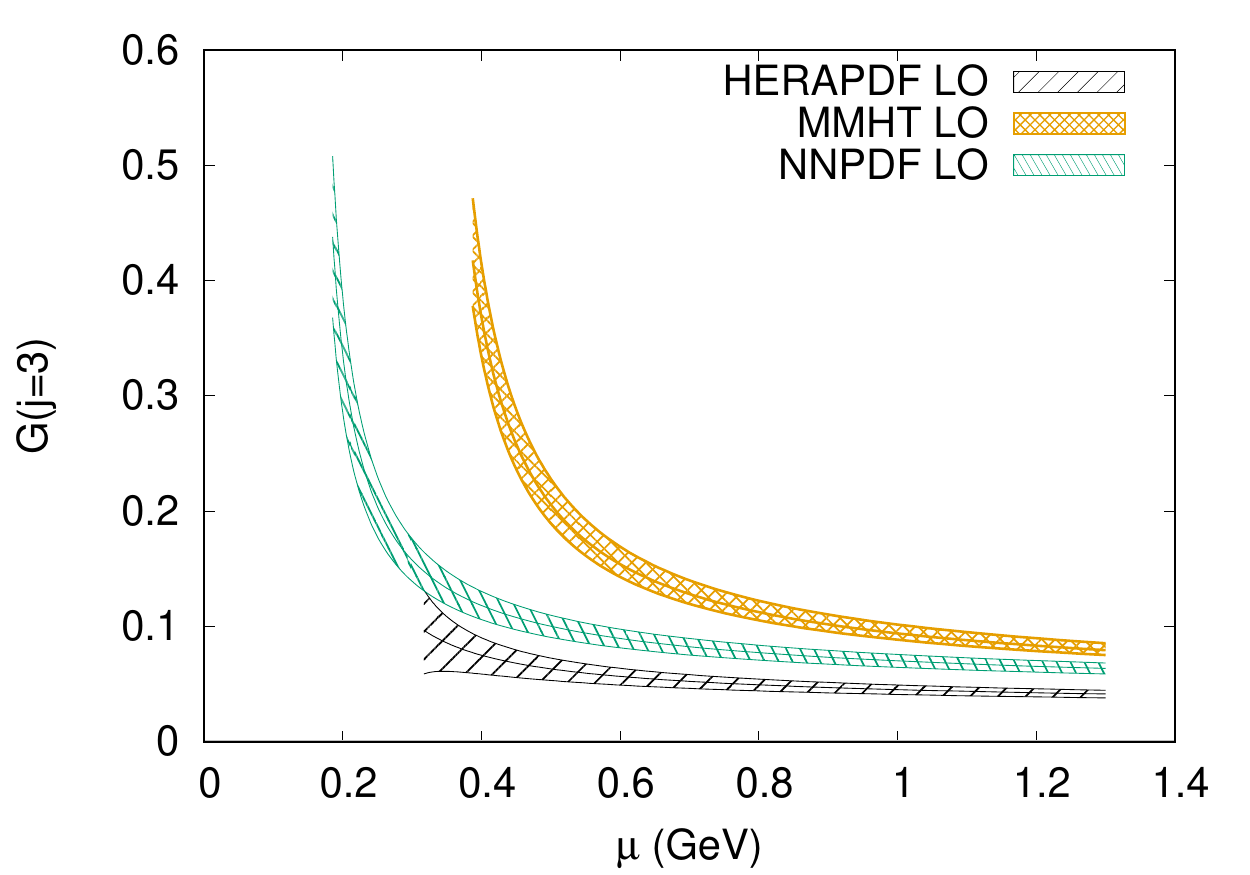}}
\hfill
\subfigure[$\Qbar(3)$ at LO]{\includegraphics[width=0.49\textwidth,trim=25 0 0 0,clip]{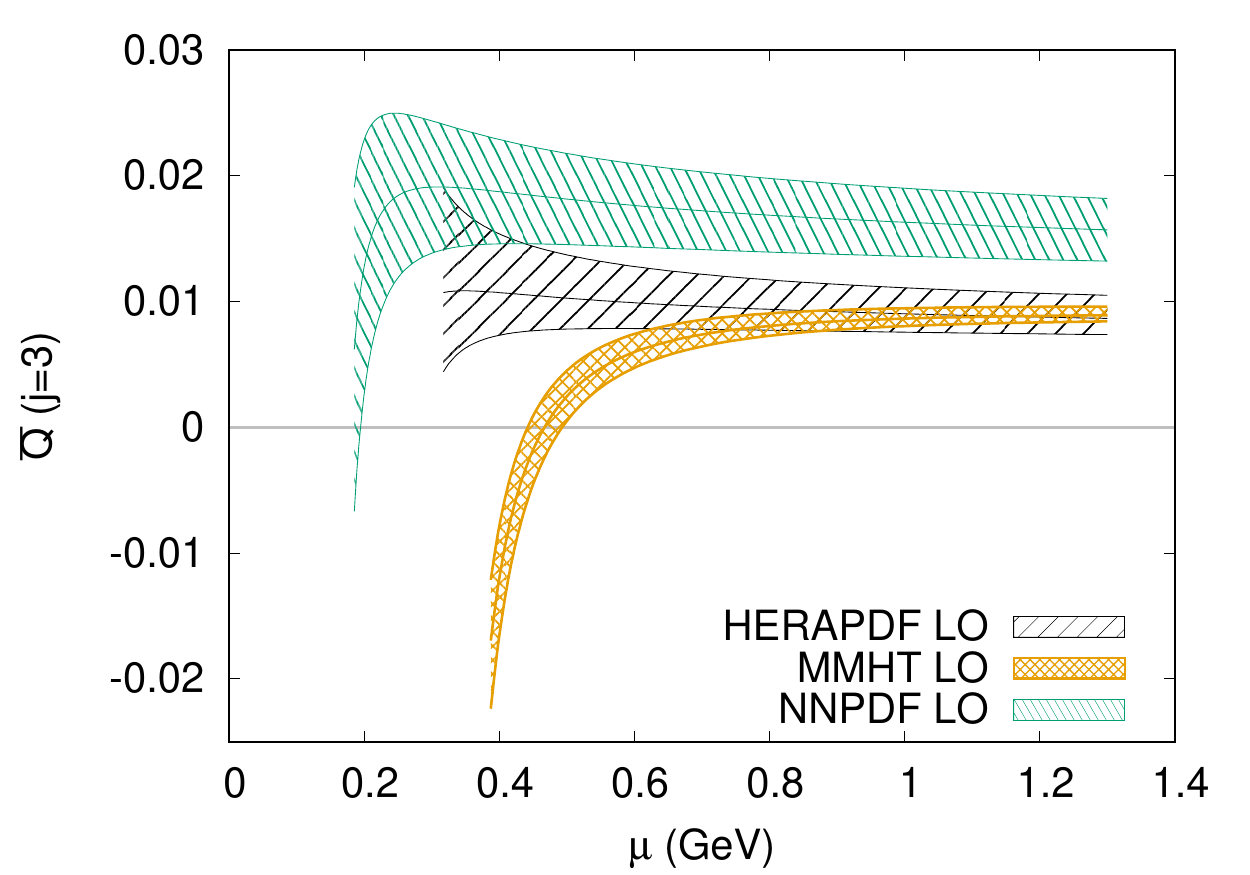}}
\\[1em]
\subfigure[$G(3)$ at NLO]{\includegraphics[width=0.49\textwidth,trim=25 0 0 0,clip]{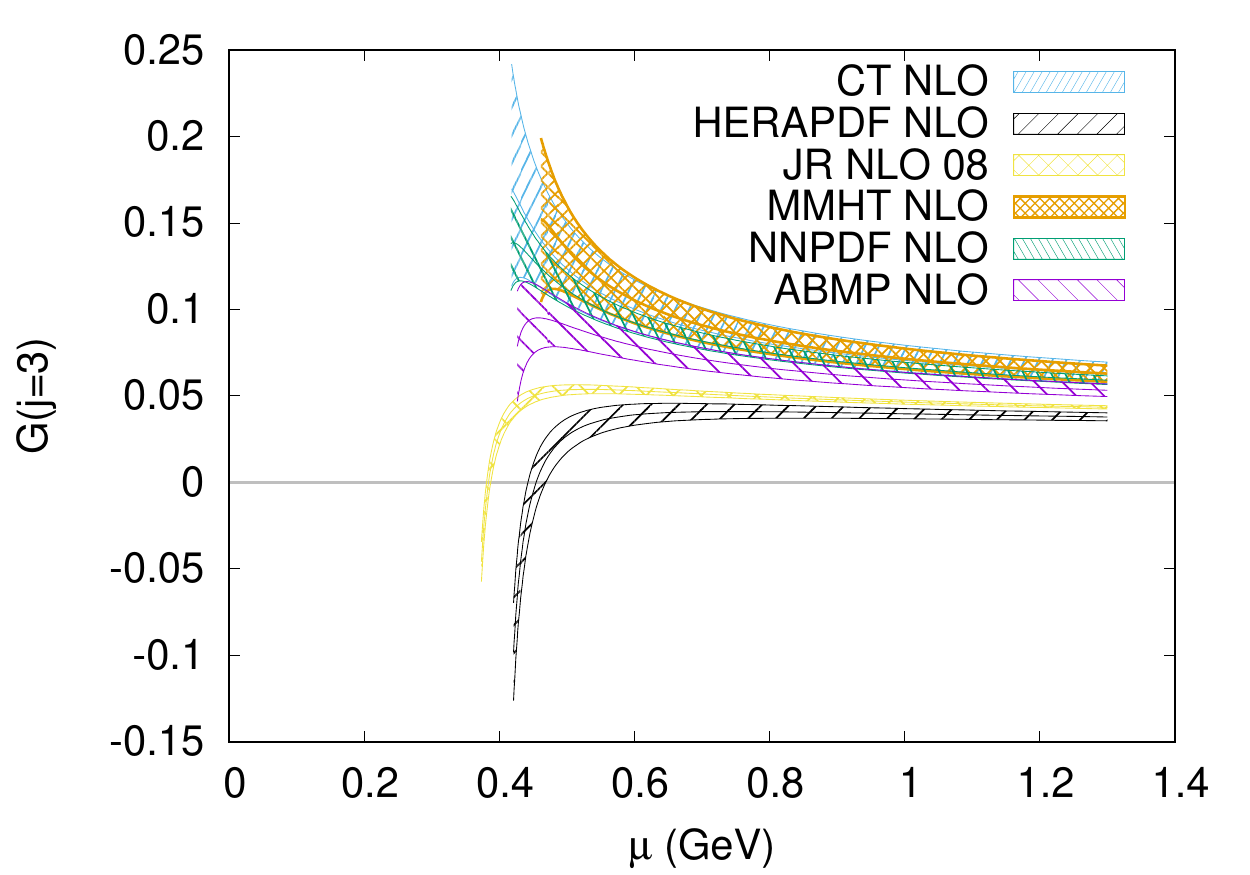}}
\hfill
\subfigure[\label{fig:mom-qbar-NLO} $\Qbar(3)$ at NLO]{\includegraphics[width=0.49\textwidth,trim=25 0 0 0,clip]{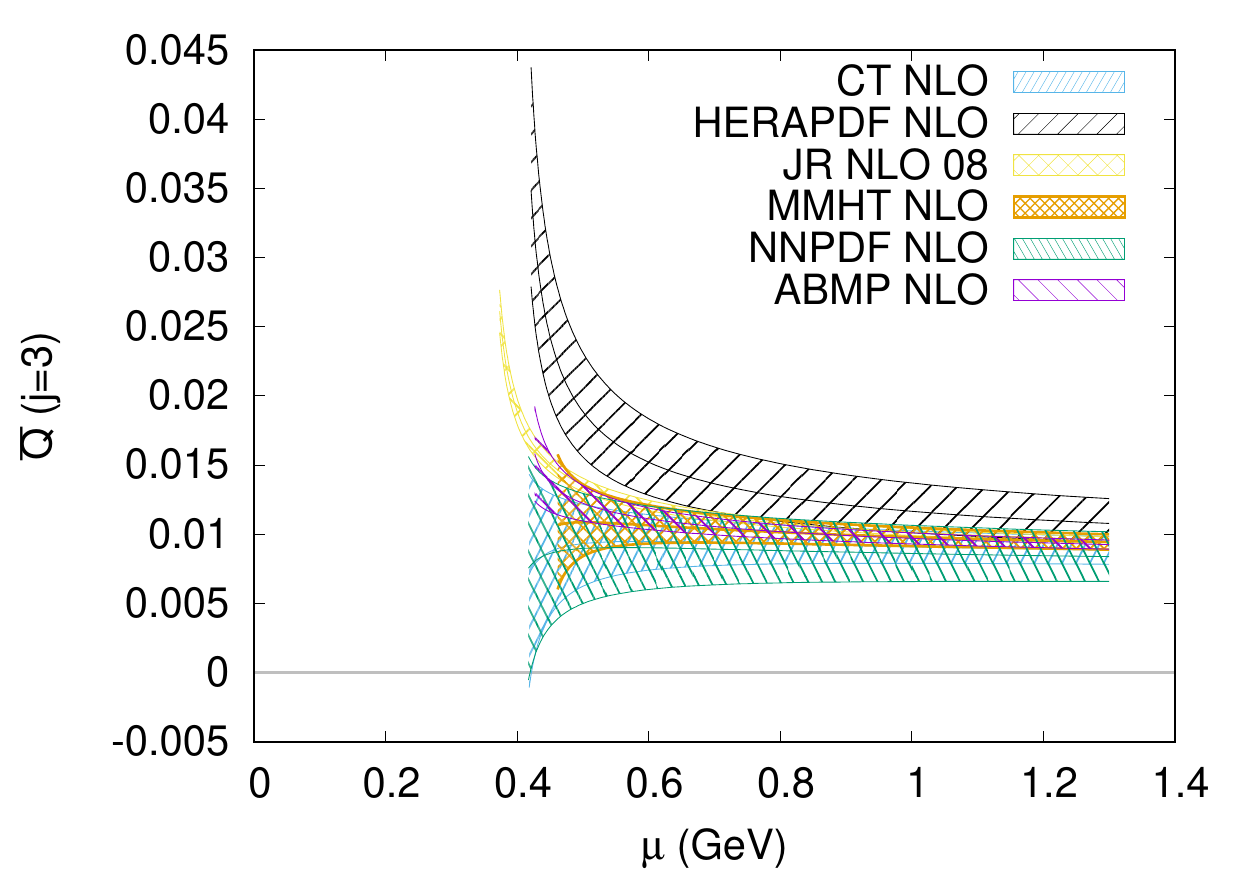}}
\\[1em]
\subfigure[$G(3)$ at NNLO]{\includegraphics[width=0.49\textwidth,trim=25 0 0 0,clip]{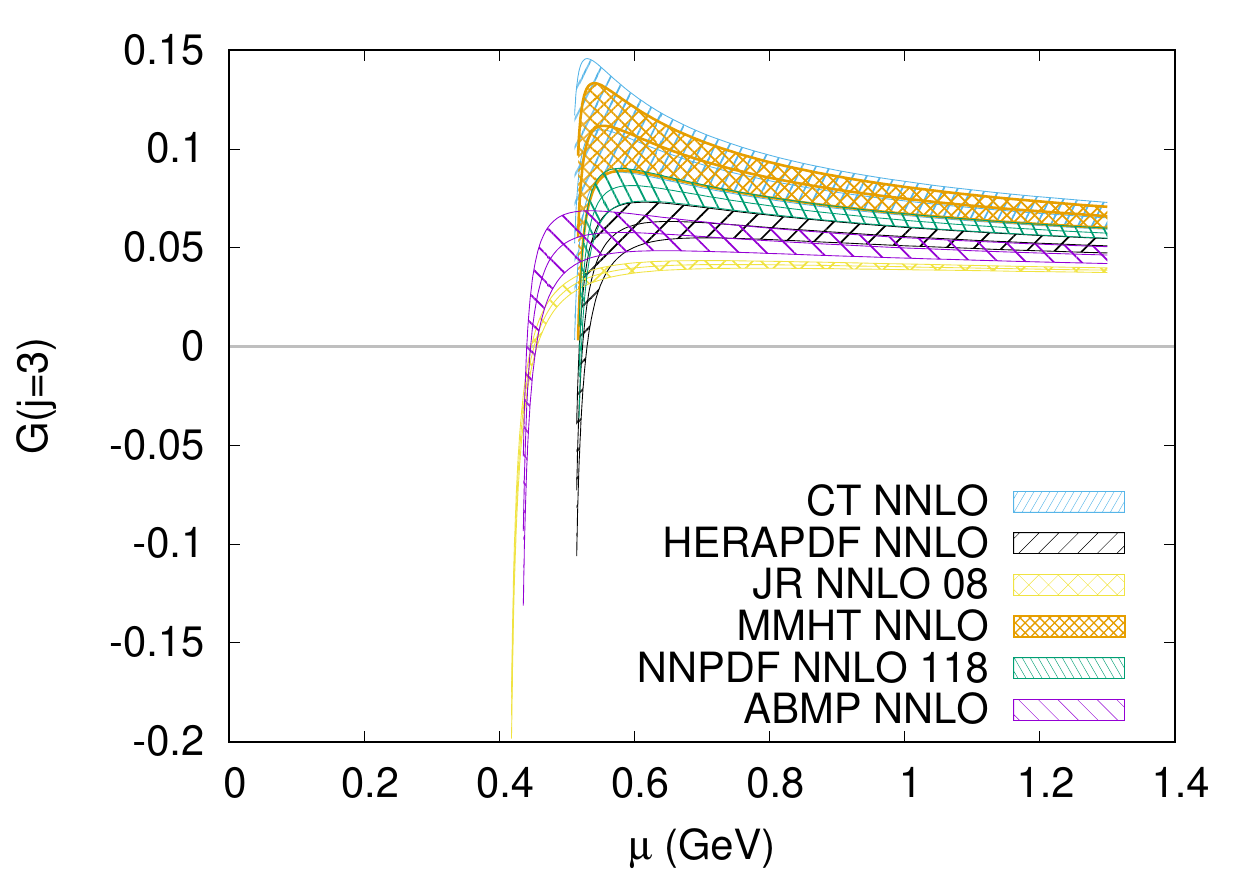}}
\hfill
\subfigure[\label{fig:mom-qbar-NNLO} $\Qbar(3)$ at NNLO]{\includegraphics[width=0.49\textwidth,trim=25 0 0 0,clip]{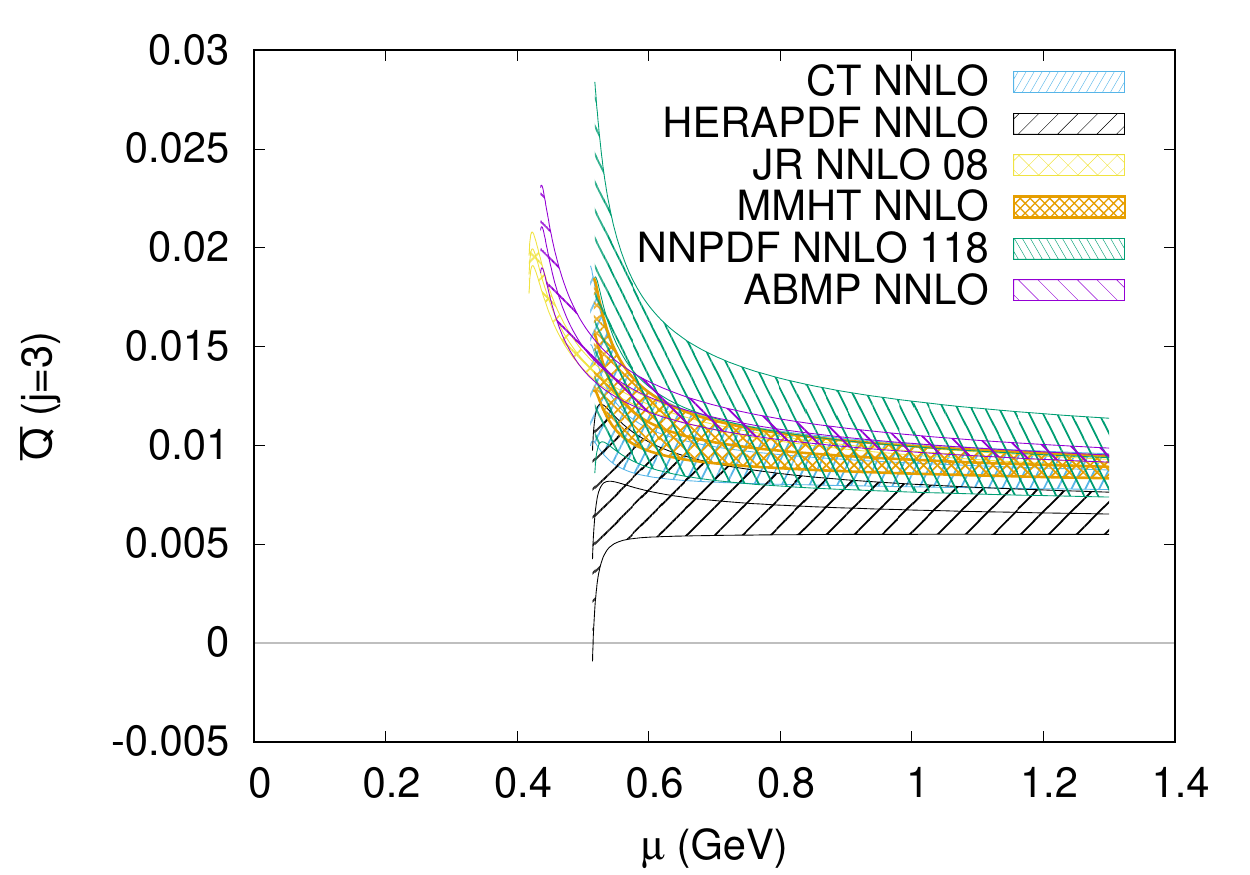}}
\end{center}
\caption{\label{fig:moments-sets} Comparison of $j=3$ moments for gluons and antiquarks between different PDF sets.  In panels (e) and (f) we have omitted the sets JR NNLO 20, NNPDF NNLO 116, and NNPDF NNLO 120 for the sake of clarity; their values lie in-between the values of other sets.}
\end{figure*}

We now discuss how given Mellin moments compare between different sets.  In addition to the parametric PDF errors provided by each set, the difference between sets may be taken as an indication of how well a given Mellin moment is known.  The situation is shown in \fig{\ref{fig:moments-sets-LO}} for $j=2$ at LO and in \fig{\ref{fig:moments-sets}} for $j=3$ at all orders.  In this case, we plot moments against $\mu$ rather than $\as$, because different PDFs sets have different values of $\as(\mu)$ at a given scale $\mu$.

Starting with the LO sets, we find that $G(j)$ evolves to a zero value and then becomes negative for $j=1.5$ and $2$.  For $j=2.5$ it either increases or decreases with $\alpha_s$, depending on the set, and for $j=3$ it increases or remains flat.  The moments $\Qbar(j)$ with $j=1.5$ and $2$ increase with $\as$.  For $j=2.5$ and $3$, the antiquark moments evolve to zero for MMHT, whereas for HERAPDF and NNPDF they decrease only slightly or remain flat.  The behaviour of $\Qbar(2)$ and $G(3)$ in \figs{\ref{fig:mom-qbar-LO}} and \ref{fig:mom-glu-LO} alone implies that there is no LO set in which either the gluon or the flavour sum of antiquark distributions evolves to zero at some low scale.  This remains true if we include the CJ LO set in our considerations (see our remark on that set in \sect{\ref{sec:pdfs}}).

Moving on to the NLO sets, we find that $G(j)$ has a zero crossing for $j=1.5$ and $2$ in all sets, whereas for $j=2.5$ and $j=3$ this only happens for some of the sets.  As was the case for LO, the moments $\Qbar(j)$ with $j=1.5$ and $2$ increase with $\as$.  We find no zero crossings for $\Qbar(2.5)$, whereas in some PDF sets the error bands of $\Qbar(3)$ just touch zero at the lowest scales, as seen in \fig{\ref{fig:mom-qbar-NLO}}.  None of the NLO sets is hence compatible with the flavour sum of antiquark distributions vanishing at low scales.

Turning to the NNLO sets, we observe that in all sets all moments $G(j)$ decrease with $\as$, either crossing zero, or coming close to zero, or being consistent with zero within their error bands.  The antiquark moments with $j=1.5$, $2$, and $2.5$ decrease with $\as$, with some of them going down to zero and others not.  For the moment $\Qbar(3)$, only HERAPDF is consistent with zero at very low scale, while the other sets are not, as seen in \fig{\ref{fig:mom-qbar-NNLO}}.  Whether any NNLO set admits the gluon or antiquark distributions to evolve to zero is investigated more closely in the next subsection.

%%%%%%%%%%%%%%%%%%%%%%%%%%%%%%%%%%%%%%%%%

\subsection{Vanishing gluon or antiquark moments}
\label{sec:zero-search}

If $g(x,\mu)$ or the sum $\bar{u}(x,\mu) + \bar{d}(x,\mu) + \bar{s}(x,\mu)$ is zero at some scale $\mu$, then all Mellin moments of these distributions must vanish at that scale, or equivalently at the corresponding value of $\as(\mu)$.  In order to test this hypothesis, we compare the different moments $G(j, \as)$ and $\Qbar(j, \as)$ for each PDF set in turn.

\begin{figure*}
\begin{center}
\subfigure[$G(j,\as) / G\bigl( j,\as(\mu_0) \bigr)$ for ABMP]{\includegraphics[width=0.49\textwidth]{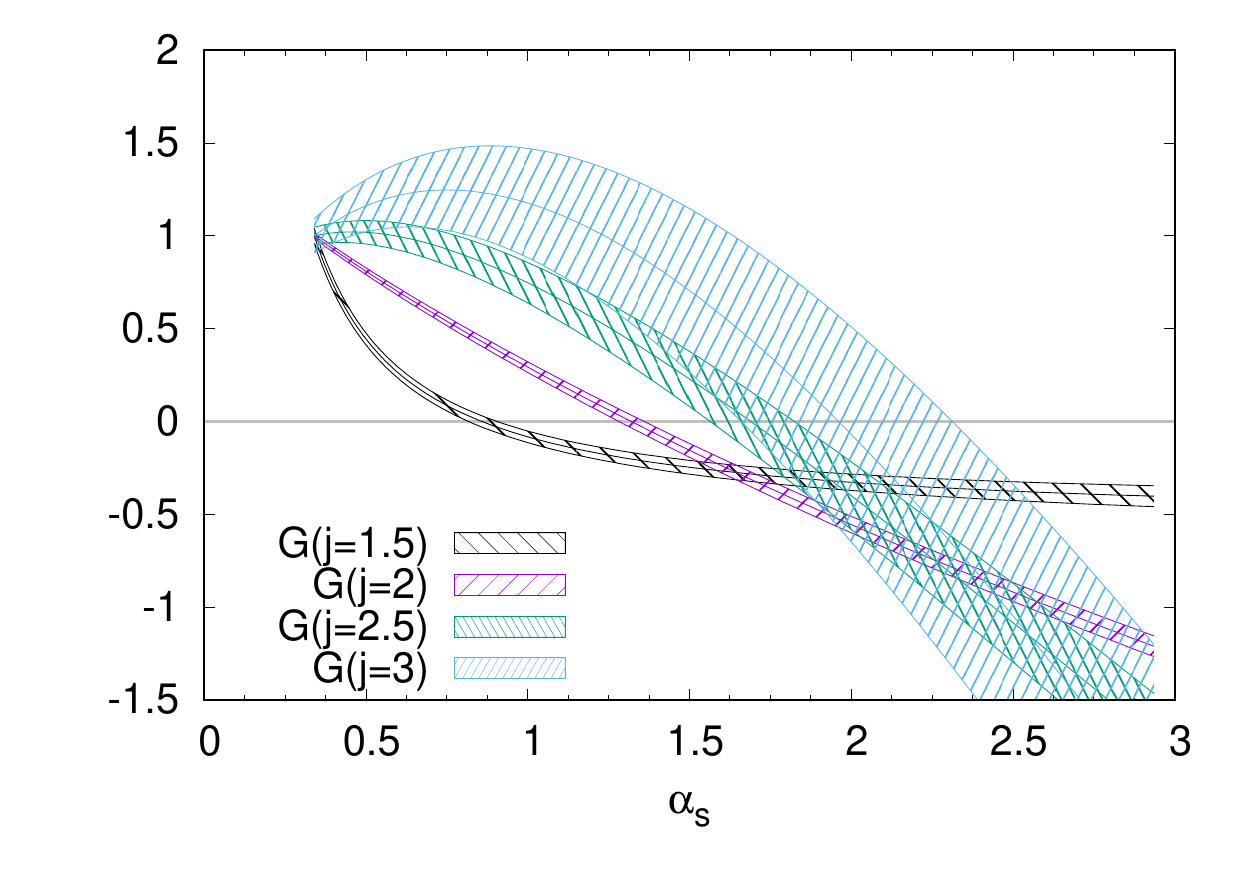}}
\hfill
\subfigure[$G(j,\as) / G\bigl( j,\as(\mu_0) \bigr)$ for JR NNLO 08]{\includegraphics[width=0.49\textwidth]{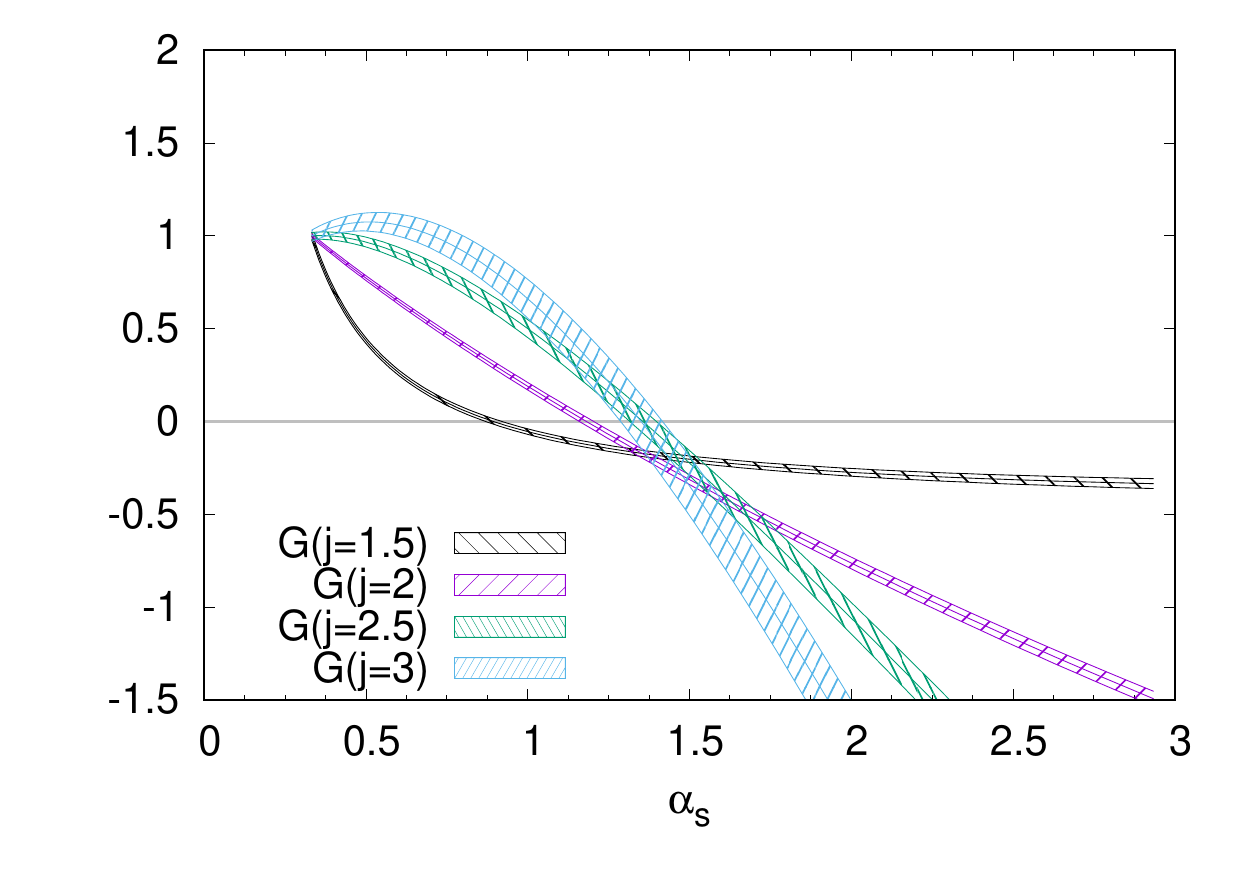}}
\\[1em]
\subfigure[\label{fig:glu-HERAPDF} $G(j,\as) / G\bigl( j,\as(\mu_0) \bigr)$ for HERAPDF NNLO]{\includegraphics[width=0.49\textwidth]{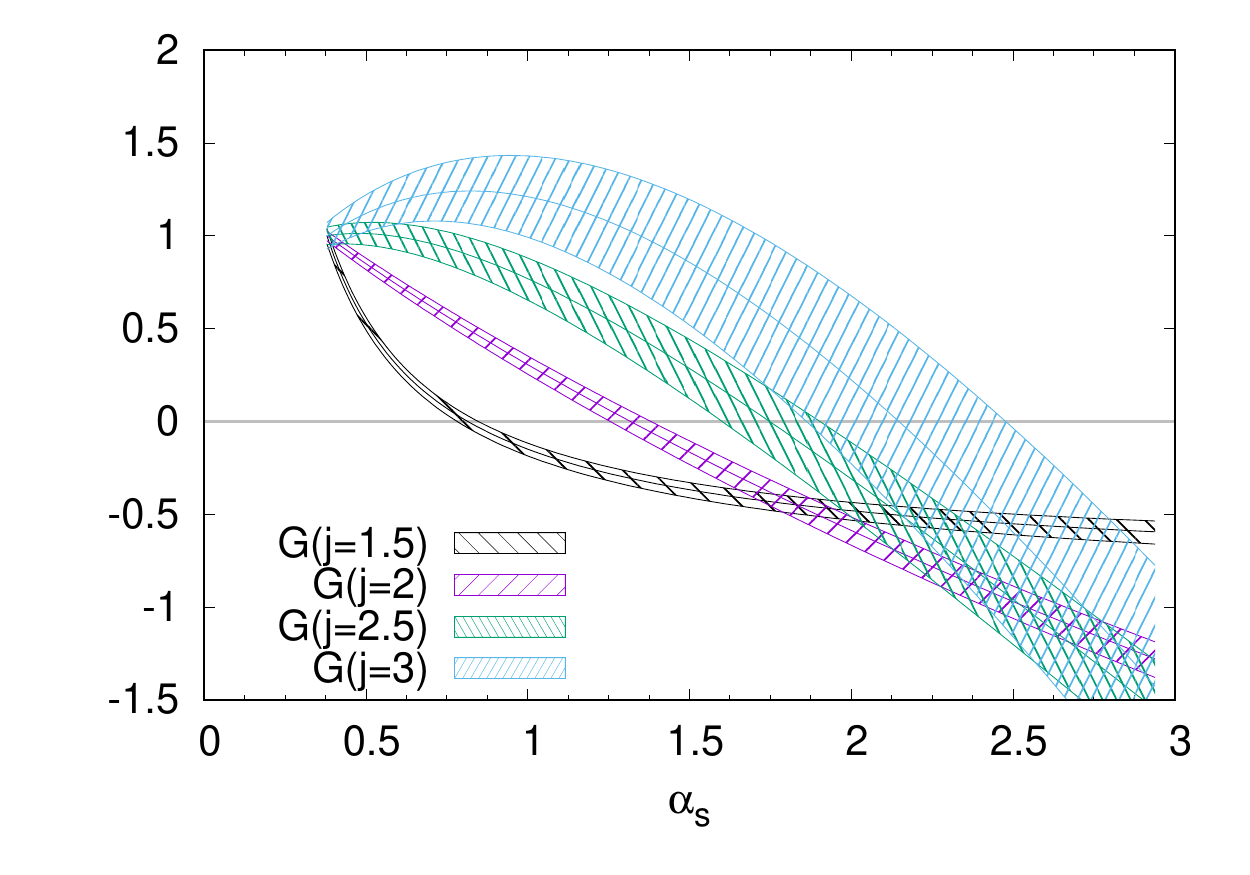}}
\hfill
\subfigure[\label{fig:glu-NNPDF118} $G(j,\as) / G\bigl( j,\as(\mu_0) \bigr)$ for NNPDF NNLO 118]{\includegraphics[width=0.49\textwidth]{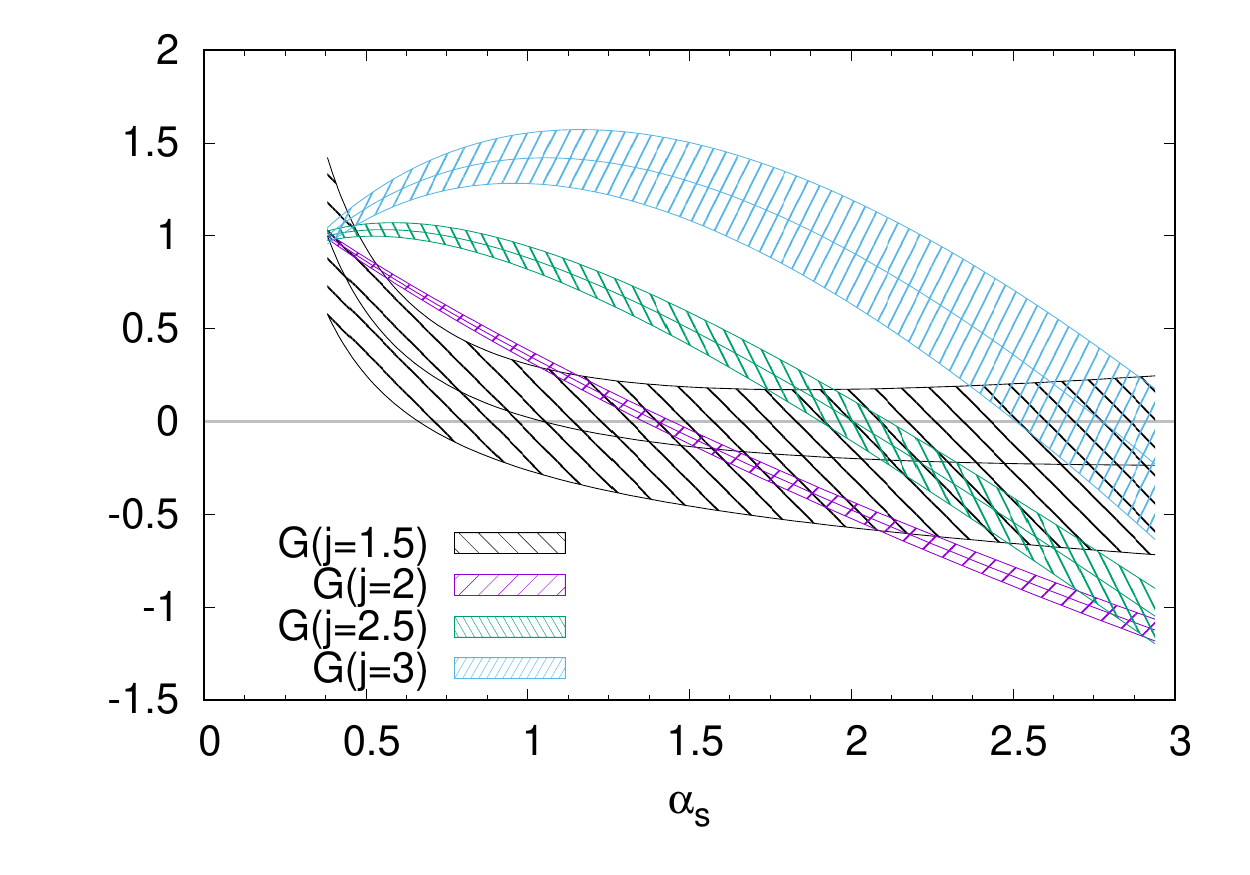}}
\end{center}
\caption{\label{fig:zeroes-gluon} Comparison of gluon moments with different $j$ for selected PDF sets.  Each moment is normalised to its value at the input scale $\mu_0 = 1.3 \gev$.}
\end{figure*}

Starting with the gluon, we see in \fig{\ref{fig:zeroes-gluon}} that for ABMP, JR NNLO 08, and HERAPDF NNLO, all moments $G(j)$ have a zero crossing but are never consistent with zero at the same value of $\as$.  In particular, the lowest moment $G(1.5)$ has already evolved to negative values at $\as$ values where the higher moments become zero, so that $g(x)$ must be negative in some $x$ range.  The situation is similar for HERAPDF NLO, for all JR sets, and for NNPDF NNLO 116.  In the 118 and 120 variants of NNPDF NNLO, the gluon moments with $j \ge 2$ have zero crossings at well separated $\as$ values, whilst $G(1.5)$ is consistent with zero over a large $\as$ range, as is shown in \fig{\ref{fig:glu-NNPDF118}} for the 118 set.  In all other PDF sets (including CJ), we find that $G(3)$ stays positive (for MMHT NNLO, the error band just touches zero at $\as \approx 3$).  We thus find no set that is compatible with a vanishing (or very small) gluon distribution at low scales.
We recall from \sect{\ref{sec:moments-x}} that for some PDF sets, the truncation uncertainty on $G(1.5)$ at the starting scale is not negligible.  This does not weaken the conclusion just stated, which for the sets in question already follows from the three moments $G(j)$ with $j \ge 2$.

Let us also note that for some scale below $\mu = 0.8 \gev$ (for NNLO) or $0.7 \gev$ (for NLO and LO), we find either that $G(1.5)$ has a zero crossing whilst all higher moments are positive, or that $G(1.5)$ can have either sign within its errors.  In the latter case, which happens for NNPDF LO, NNPDF NNLO 118 and 120, and CT NLO), no strong conclusions can be drawn.  The former case indicates that $g(x)$ is negative at low $x$ and positive at high $x$, which implies that its probability interpretation is lost.

\begin{figure*}
\begin{center}
\subfigure[\label{fig:qbar-HERA-NNLO} $\Qbar(j,\as) /
\Qbar\bigl( j,\as(\mu_0) \bigr)$ for HERAPDF NNLO]{\includegraphics[width=0.49\textwidth]{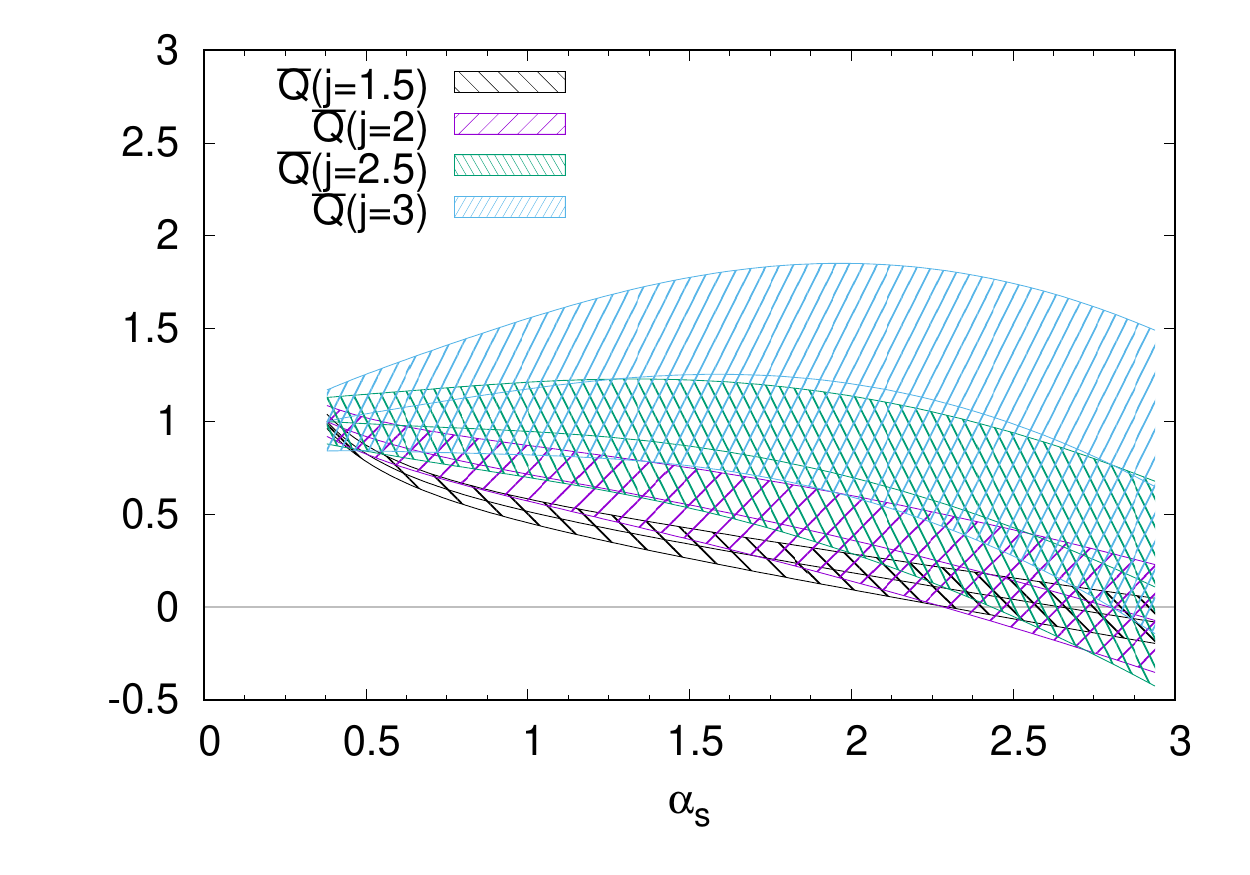}}
\hfill
\subfigure[\label{fig:qbar-HERA-NLO} $\Qbar(j,\as) / \Qbar\bigl( j,\as(\mu_0) \bigr)$ for HERAPDF NLO]{\includegraphics[width=0.49\textwidth]{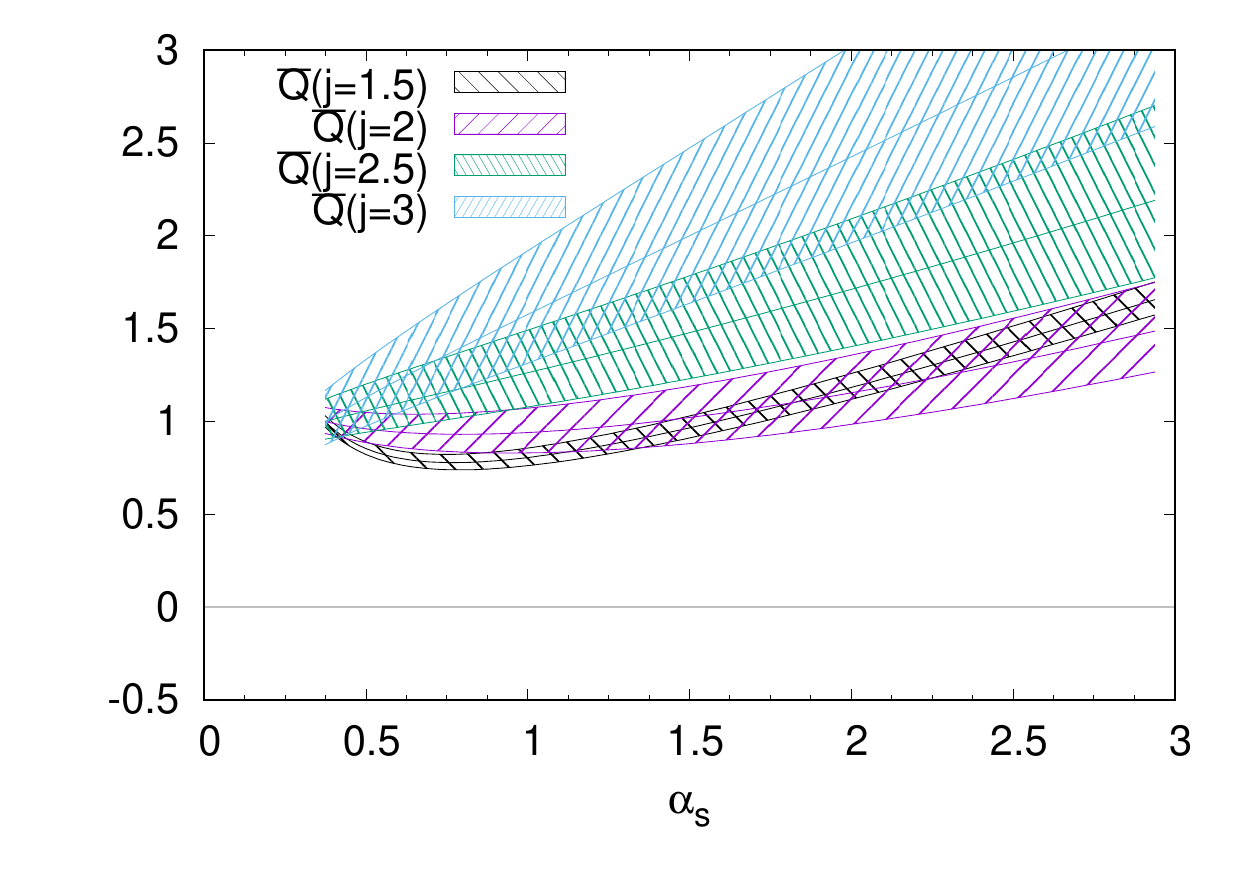}}
\end{center}
\caption{\label{fig:zeroes-qbar} Comparison of antiquark moments with different $j$ for selected PDF sets.  Each moment is normalised to its value at the input scale $\mu_0 = 1.3 \gev$.}
\end{figure*}

We now turn to antiquarks and note that, in general, a moment $\Qbar(j)$ is positive at the scale where $G(j)$ with the same $j$ has a zero crossing.  To find zeroes of $\Qbar(j)$, one thus has to go to yet lower scales.  The only set in which all antiquark moments are consistent with zero at the same scale is HERAPDF NNLO, shown in \fig{\ref{fig:qbar-HERA-NNLO}}.  However, this happens at $\as$ values above $2.8$ (corresponding to $\mu \approx 0.51 \gev$).  At these $\as$ values, all gluon moments of the same set are negative, as seen in \fig{\ref{fig:glu-HERAPDF}}.  The probability interpretation of PDFs is therefore lost at these low scales, and the PDF set is not consistent with a scenario in which at some scale antiquark distributions vanish whilst gluon and quark distributions are positive.

In all other PDF sets, there is at least one moment $\Qbar(j)$ that remains significantly different from zero over the full $\as$ range we consider.  For the NNLO sets other than HERAPDF, this is the $j=3$ moment, whereas for the NLO and LO sets (including CJ) it is the moment with $j=2$.  For some sets, all antiquark moments remain positive, as in the example of \fig{\ref{fig:qbar-HERA-NLO}}.

\section{Conclusions}
\label{sec:sum}

In this work we investigate the possibility that the gluon or the antiquark PDFs in the proton go to zero when evolved to low scales, using the DGLAP equations with splitting functions computed in perturbation theory.  To do so, we take a variety of current PDF sets and compute Mellin moments of PDFs for $j=1.5$, $2$, $2.5$, and $3$ at a reference scale $\mu_0 = 1.3 \gev$.  We then evolve these moments down to lower scales, stopping at the point where $\as(\mu) = 3$.  Our comparison of these moments at LO, NLO, and NNLO indicates that perturbation theory ceases to converge at much lower values of $\as$, so that considering even larger values would make little sense.

In several NLO and NNLO PDF sets, the gluon moments for all considered $j$ go to zero under evolution, but they do so at different scales $\mu$.  No PDF set is found to be compatible with the gluon PDF vanishing at any scale.  We note that the $j=1.5$ moment eventually evolves to negative values in almost all sets at some scale $\mu$ below $0.8 \gev$, which indicates that $g(x) < 0$ at low $x$ and hence cannot be interpreted as a probability density.

For antiquarks, we take the flavour sum $\bar{u}(x) + \bar{d}(x) + \bar{s}(x)$ and find that at least some of its Mellin moments remain positive up to the largest $\as$ we consider.  An exception to this is the HERAPDF NNLO set, for which all antiquark moments congregate around zero for $\as \gsim 2.8$.  At these scales, however, the gluon density in the same set has negative Mellin moments, so that we cannot interpret the PDFs as densities.

Our findings are fully consistent with those of the Dortmund group \cite{Gluck:1989ze,Gluck:1991ng,Gluck:1994uf,Jimenez-Delgado:2014twa}, who concluded from their fits that PDFs that describe high-energy scattering data cannot be generated by perturbative radiation from an input that involves only valence quark densities at some low momentum scale.  We think that our study is a valuable complement to the Dortmund approach, in particular because we use the results of a broad array of PDF fits, which differ in their data selection, parameterisation of PDFs, as well as the details of the theory description of hard cross sections.

Which scenarios does this situation leave for connecting the PDFs determined from data at high scales with PDFs computed in models at low scales?  We see several obvious possibilities.  The least dramatic one would be to consider PDFs defined in a perturbative renormalisation scheme other than $\overline{\text{MS}}$ and to identify these PDFs with the results of low-energy computations.  Perhaps more plausible is that the evolution of PDFs is modified by non-perturbative effects at low scales, as has been argued for the case of the running coupling~\cite{Deur:2016tte}.  This is also suggested by our comparison of Mellin moments evolved at different orders, and by the study \cite{Altenbuchinger:2010sz} of evolution in the polarised parton sector.  Finally, it may well be that even at the lowest scales one can sensibly consider, the parton content of the proton is not limited to just ``valence quarks'' but involves antiquarks or gluons or both, as several low-energy models suggest.

%%%%%%%%%%%%%%%%%%%%%%%%%%%%%%%

\section*{Acknowledgements}

It is a pleasure to thank Andreas Sch\"afer for valuable remarks on the manuscript.

\section*{Note added in proof}

After this work was completed, we were made aware of the paper \cite{RuizArriola:1998er}.  Using PDF sets available at the time, that study evolved PDFs in $x$ space down to low scales.  In particular, parton distributions were found to become negative below $\mu \sim 600 \mev$.  The qualitative conclusions of \cite{RuizArriola:1998er} agree with those of our work, disfavouring the idea that PDFs at high scales can be obtained from DGLAP evolution of PDFs computed in quark models.

%%%%%%%%%%%%%%%%%%%%%%%%%%%%%%%

\phantomsection
\addcontentsline{toc}{section}{References}

\bibliographystyle{JHEP}
\bibliography{evolve}

\providecommand{\href}[2]{#2}\begingroup\raggedright\begin{thebibliography}{10}

\bibitem{Tanabashi:2018oca}
{\scshape Particle Data Group}, M.~Tanabashi et~al.,
  \emph{{Review of Particle Physics}},
  \href{https://doi.org/10.1103/PhysRevD.98.030001}{\emph{Phys. Rev.}
  {\bfseries D98} (2018) 030001}.

\bibitem{Parisi:1976fz}
G.~Parisi and R.~Petronzio, \emph{{On the Breaking of Bjorken Scaling}},
  \href{https://doi.org/10.1016/0370-2693(76)90088-5}{\emph{Phys. Lett.}
  {\bfseries 62B} (1976) 331}.

\bibitem{Novikov:1976dd}
V.~A. Novikov, M.~A. Shifman, A.~I. Vainshtein and V.~I. Zakharov, \emph{{Naive
  Quark Model and Deep Inelastic Scattering}},
  \href{https://doi.org/10.1016/0003-4916(77)90241-X}{\emph{Annals Phys.}
  {\bfseries 105} (1977) 276}.

\bibitem{Gluck:1977ah}
M.~Gl{\"u}ck and E.~Reya, \emph{{Dynamical Determination of Parton and Gluon
  Distributions in Quantum Chromodynamics}},
  \href{https://doi.org/10.1016/0550-3213(77)90393-5}{\emph{Nucl. Phys.}
  {\bfseries B130} (1977) 76}.

\bibitem{Gluck:1989ze}
M.~Gl{\"u}ck, E.~Reya and A.~Vogt, \emph{{Radiatively generated parton
  distributions for high-energy collisions}},
  \href{https://doi.org/10.1007/BF01572029}{\emph{Z. Phys.} {\bfseries C48}
  (1990) 471}.

\bibitem{Gluck:1991ng}
M.~Gl{\"u}ck, E.~Reya and A.~Vogt, \emph{{Parton distributions for high-energy
  collisions}}, \href{https://doi.org/10.1007/BF01483880}{\emph{Z. Phys.}
  {\bfseries C53} (1992) 127}.

\bibitem{Gluck:1994uf}
M.~Gl{\"u}ck, E.~Reya and A.~Vogt, \emph{{Dynamical parton distributions of the
  proton and small x physics}},
  \href{https://doi.org/10.1007/BF01624586}{\emph{Z. Phys.} {\bfseries C67}
  (1995) 433}.

\bibitem{Jimenez-Delgado:2014twa}
P.~Jimenez-Delgado and E.~Reya, \emph{{Delineating parton distributions and the
  strong coupling}},
  \href{https://doi.org/10.1103/PhysRevD.89.074049}{\emph{Phys. Rev.}
  {\bfseries D89} (2014) 074049}
  [\href{https://arxiv.org/abs/1403.1852}{{\ttfamily 1403.1852}}].

\bibitem{Gluck:2000dy}
M.~Gl{\"u}ck, E.~Reya, M.~Stratmann and W.~Vogelsang, \emph{{Models for the
  polarized parton distributions of the nucleon}},
  \href{https://doi.org/10.1103/PhysRevD.63.094005}{\emph{Phys. Rev.}
  {\bfseries D63} (2001) 094005}
  [\href{https://arxiv.org/abs/hep-ph/0011215}{{\ttfamily hep-ph/0011215}}].

\bibitem{Jaffe:1980ti}
R.~L. Jaffe and G.~G. Ross, \emph{{Normalizing the Renormalization Group
  Analysis of Deep Inelastic Leptoproduction}},
  \href{https://doi.org/10.1016/0370-2693(80)90521-3}{\emph{Phys. Lett.}
  {\bfseries 93B} (1980) 313}.

\bibitem{Traini:1997jz}
M.~Traini, A.~Mair, A.~Zambarda and V.~Vento, \emph{{Constituent quarks and
  parton distributions}},
  \href{https://doi.org/10.1016/S0375-9474(96)00450-2}{\emph{Nucl. Phys.}
  {\bfseries A614} (1997) 472}.

\bibitem{Scopetta:2002xq}
S.~Scopetta and V.~Vento, \emph{{Generalized parton distributions in
  constituent quark models}},
  \href{https://doi.org/10.1140/epja/i2002-10120-y}{\emph{Eur. Phys. J.}
  {\bfseries A16} (2003) 527}
  [\href{https://arxiv.org/abs/hep-ph/0201265}{{\ttfamily hep-ph/0201265}}].

\bibitem{Rinaldi:2014ddl}
M.~Rinaldi, S.~Scopetta, M.~Traini and V.~Vento, \emph{{Double parton
  correlations and constituent quark models: a Light Front approach to the
  valence sector}}, \href{https://doi.org/10.1007/JHEP12(2014)028}{\emph{JHEP}
  {\bfseries 12} (2014) 028} [\href{https://arxiv.org/abs/1409.1500}{{\ttfamily
  1409.1500}}].

\bibitem{Altarelli:1973ff}
G.~Altarelli, N.~Cabibbo, L.~Maiani and R.~Petronzio, \emph{{The Nucleon as a
  bound state of three quarks and deep inelastic phenomena}},
  \href{https://doi.org/10.1016/0550-3213(74)90452-0}{\emph{Nucl. Phys.}
  {\bfseries B69} (1974) 531}.

\bibitem{Scopetta:1997wk}
S.~Scopetta, V.~Vento and M.~Traini, \emph{{Towards a unified picture of
  constituent and current quarks}},
  \href{https://doi.org/10.1016/S0370-2693(97)01599-2}{\emph{Phys. Lett.}
  {\bfseries B421} (1998) 64}
  [\href{https://arxiv.org/abs/hep-ph/9708262}{{\ttfamily hep-ph/9708262}}].

\bibitem{Noguera:2004jq}
S.~Noguera, S.~Scopetta and V.~Vento, \emph{{Relativity and constituent quark
  structure in model calculations of parton distributions}},
  \href{https://doi.org/10.1103/PhysRevD.70.094018}{\emph{Phys. Rev.}
  {\bfseries D70} (2004) 094018}
  [\href{https://arxiv.org/abs/hep-ph/0409059}{{\ttfamily hep-ph/0409059}}].

\bibitem{Thomas:1983fh}
A.~W. Thomas, \emph{{A Limit on the Pionic Component of the Nucleon Through
  SU(3) Flavor Breaking in the Sea}},
  \href{https://doi.org/10.1016/0370-2693(83)90026-6}{\emph{Phys. Lett.}
  {\bfseries 126B} (1983) 97}.

\bibitem{Kumano:1997cy}
S.~Kumano, \emph{{Flavor asymmetry of anti-quark distributions in the
  nucleon}}, \href{https://doi.org/10.1016/S0370-1573(98)00016-7}{\emph{Phys.
  Rept.} {\bfseries 303} (1998) 183}
  [\href{https://arxiv.org/abs/hep-ph/9702367}{{\ttfamily hep-ph/9702367}}].

\bibitem{Traini:2013zqa}
M.~Traini, \emph{{Next-to-next-to-leading-order nucleon parton distributions
  from a light-cone quark model dressed with its virtual meson cloud}},
  \href{https://doi.org/10.1103/PhysRevD.89.034021}{\emph{Phys. Rev.}
  {\bfseries D89} (2014) 034021}
  [\href{https://arxiv.org/abs/1309.5814}{{\ttfamily 1309.5814}}].

\bibitem{Wang:2016ndh}
X.~G. Wang, C.-R. Ji, W.~Melnitchouk, Y.~Salamu, A.~W. Thomas and P.~Wang,
  \emph{{Strange quark asymmetry in the proton in chiral effective theory}},
  \href{https://doi.org/10.1103/PhysRevD.94.094035}{\emph{Phys. Rev.}
  {\bfseries D94} (2016) 094035}
  [\href{https://arxiv.org/abs/1610.03333}{{\ttfamily 1610.03333}}].

\bibitem{Kofler:2017uzq}
S.~Kofler and B.~Pasquini, \emph{{Collinear parton distributions and the
  structure of the nucleon sea in a light-front meson-cloud model}},
  \href{https://doi.org/10.1103/PhysRevD.95.094015}{\emph{Phys. Rev.}
  {\bfseries D95} (2017) 094015}
  [\href{https://arxiv.org/abs/1701.07839}{{\ttfamily 1701.07839}}].

\bibitem{Koepf:1995yh}
W.~Koepf, L.~L. Frankfurt and M.~Strikman, \emph{{The Nucleon's virtual meson
  cloud and deep inelastic lepton scattering}},
  \href{https://doi.org/10.1103/PhysRevD.53.2586}{\emph{Phys. Rev.} {\bfseries
  D53} (1996) 2586} [\href{https://arxiv.org/abs/hep-ph/9507218}{{\ttfamily
  hep-ph/9507218}}].

\bibitem{Strikman:2003gz}
M.~Strikman and C.~Weiss, \emph{{Chiral dynamics and the growth of the
  nucleon's gluonic transverse size at small x}},
  \href{https://doi.org/10.1103/PhysRevD.69.054012}{\emph{Phys. Rev.}
  {\bfseries D69} (2004) 054012}
  [\href{https://arxiv.org/abs/hep-ph/0308191}{{\ttfamily hep-ph/0308191}}].

\bibitem{Diakonov:1996sr}
D.~Diakonov, V.~Petrov, P.~Pobylitsa, M.~V. Polyakov and C.~Weiss,
  \emph{{Nucleon parton distributions at low normalization point in the large
  $N_c$ limit}},
  \href{https://doi.org/10.1016/S0550-3213(96)00486-5}{\emph{Nucl. Phys.}
  {\bfseries B480} (1996) 341}
  [\href{https://arxiv.org/abs/hep-ph/9606314}{{\ttfamily hep-ph/9606314}}].

\bibitem{Diakonov:1997vc}
D.~Diakonov, V.~{\relax Yu}. Petrov, P.~V. Pobylitsa, M.~V. Polyakov and
  C.~Weiss, \emph{{Unpolarized and polarized quark distributions in the large
  N(c) limit}}, \href{https://doi.org/10.1103/PhysRevD.56.4069}{\emph{Phys.
  Rev.} {\bfseries D56} (1997) 4069}
  [\href{https://arxiv.org/abs/hep-ph/9703420}{{\ttfamily hep-ph/9703420}}].

\bibitem{Lampe:1998eu}
B.~Lampe and E.~Reya, \emph{{Spin physics and polarized structure functions}},
  \href{https://doi.org/10.1016/S0370-1573(99)00100-3}{\emph{Phys. Rept.}
  {\bfseries 332} (2000) 1}
  [\href{https://arxiv.org/abs/hep-ph/9810270}{{\ttfamily hep-ph/9810270}}].

\bibitem{Thomas:2008ga}
A.~W. Thomas, \emph{{Interplay of Spin and Orbital Angular Momentum in the
  Proton}}, \href{https://doi.org/10.1103/PhysRevLett.101.102003}{\emph{Phys.
  Rev. Lett.} {\bfseries 101} (2008) 102003}
  [\href{https://arxiv.org/abs/0803.2775}{{\ttfamily 0803.2775}}].

\bibitem{Altenbuchinger:2010sz}
M.~Altenbuchinger, P.~H{\"a}gler, W.~Weise and E.~M. Henley, \emph{{Spin
  structure of the nucleon: QCD evolution, lattice results and models}},
  \href{https://doi.org/10.1140/epja/i2011-11140-2}{\emph{Eur. Phys. J.}
  {\bfseries A47} (2011) 140}
  [\href{https://arxiv.org/abs/1012.4409}{{\ttfamily 1012.4409}}].

\bibitem{deFlorian:2019egz}
D.~de~Florian and W.~Vogelsang, \emph{{Spin budget of the proton at NNLO and
  beyond}}, \href{https://doi.org/10.1103/PhysRevD.99.054001}{\emph{Phys. Rev.}
  {\bfseries D99} (2019) 054001}
  [\href{https://arxiv.org/abs/1902.04636}{{\ttfamily 1902.04636}}].

\bibitem{Deur:2016tte}
A.~Deur, S.~J. Brodsky and G.~F. de~Teramond, \emph{{The QCD Running
  Coupling}}, \href{https://doi.org/10.1016/j.ppnp.2016.04.003}{\emph{Prog.
  Part. Nucl. Phys.} {\bfseries 90} (2016) 1}
  [\href{https://arxiv.org/abs/1604.08082}{{\ttfamily 1604.08082}}].

\bibitem{DallaBrida:2016kgh}
{\scshape ALPHA} collaboration, M.~Dalla~Brida, P.~Fritzsch, T.~Korzec,
  A.~Ramos, S.~Sint and R.~Sommer, \emph{{Slow running of the Gradient Flow
  coupling from 200 MeV to 4 GeV in $N_{\rm f}=3$ QCD}},
  \href{https://doi.org/10.1103/PhysRevD.95.014507}{\emph{Phys. Rev.}
  {\bfseries D95} (2017) 014507}
  [\href{https://arxiv.org/abs/1607.06423}{{\ttfamily 1607.06423}}].

\bibitem{Bruno:2017gxd}
{\scshape ALPHA} collaboration, M.~Bruno, M.~Dalla~Brida, P.~Fritzsch,
  T.~Korzec, A.~Ramos, S.~Schaefer et~al., \emph{{QCD Coupling from a
  Nonperturbative Determination of the Three-Flavor $\Lambda$ Parameter}},
  \href{https://doi.org/10.1103/PhysRevLett.119.102001}{\emph{Phys. Rev. Lett.}
  {\bfseries 119} (2017) 102001}
  [\href{https://arxiv.org/abs/1706.03821}{{\ttfamily 1706.03821}}].

\bibitem{Zafeiropoulos:2019flq}
S.~Zafeiropoulos, P.~Boucaud, F.~De~Soto, J.~Rodr\'{i}guez-Quintero and
  J.~Segovia, \emph{{The strong running coupling from the gauge sector of
  Domain Wall lattice QCD with physical quark masses}},
  \href{https://doi.org/10.1103/PhysRevLett.122.162002}{\emph{Phys. Rev. Lett.}
  {\bfseries 122} (2019) 162002}
  [\href{https://arxiv.org/abs/1902.08148}{{\ttfamily 1902.08148}}].

\bibitem{Ellis:1991qj}
R.~K. Ellis, W.~J. Stirling and B.~R. Webber, \emph{{QCD and collider
  physics}}, {\emph{Camb. Monogr. Part. Phys. Nucl. Phys. Cosmol.} {\bfseries
  8} (1996) 1}.

\bibitem{Moch:2004pa}
S.~Moch, J.~A.~M. Vermaseren and A.~Vogt, \emph{{The Three loop splitting
  functions in QCD: The Nonsinglet case}},
  \href{https://doi.org/10.1016/j.nuclphysb.2004.03.030}{\emph{Nucl. Phys.}
  {\bfseries B688} (2004) 101}
  [\href{https://arxiv.org/abs/hep-ph/0403192}{{\ttfamily hep-ph/0403192}}].

\bibitem{Vogt:2004mw}
A.~Vogt, S.~Moch and J.~A.~M. Vermaseren, \emph{{The Three-loop splitting
  functions in QCD: The Singlet case}},
  \href{https://doi.org/10.1016/j.nuclphysb.2004.04.024}{\emph{Nucl. Phys.}
  {\bfseries B691} (2004) 129}
  [\href{https://arxiv.org/abs/hep-ph/0404111}{{\ttfamily hep-ph/0404111}}].

\bibitem{Larin:1996wd}
S.~A. Larin, P.~Nogueira, T.~van Ritbergen and J.~A.~M. Vermaseren, \emph{{The
  Three loop QCD calculation of the moments of deep inelastic structure
  functions}}, \href{https://doi.org/10.1016/S0550-3213(97)80038-7}{\emph{Nucl.
  Phys.} {\bfseries B492} (1997) 338}
  [\href{https://arxiv.org/abs/hep-ph/9605317}{{\ttfamily hep-ph/9605317}}].

\bibitem{Moch:2017uml}
S.~Moch, B.~Ruijl, T.~Ueda, J.~A.~M. Vermaseren and A.~Vogt, \emph{{Four-Loop
  Non-Singlet Splitting Functions in the Planar Limit and Beyond}},
  \href{https://doi.org/10.1007/JHEP10(2017)041}{\emph{JHEP} {\bfseries 10}
  (2017) 041} [\href{https://arxiv.org/abs/1707.08315}{{\ttfamily
  1707.08315}}].

\bibitem{Vogt:2018miu}
A.~Vogt, F.~Herzog, S.~Moch, B.~Ruijl, T.~Ueda and J.~A.~M. Vermaseren,
  \emph{{Anomalous dimensions and splitting functions beyond the
  next-to-next-to-leading order}},
  \href{https://doi.org/10.22323/1.303.0050}{\emph{PoS} {\bfseries LL2018}
  (2018) 050} [\href{https://arxiv.org/abs/1808.08981}{{\ttfamily
  1808.08981}}].

\bibitem{Buckley:2014ana}
A.~Buckley, J.~Ferrando, S.~Lloyd, K.~Nordstr{\"o}m, B.~Page, M.~R{\"u}fenacht
  et~al., \emph{{LHAPDF6: parton density access in the LHC precision era}},
  \href{https://doi.org/10.1140/epjc/s10052-015-3318-8}{\emph{Eur. Phys. J.}
  {\bfseries C75} (2015) 132}
  [\href{https://arxiv.org/abs/1412.7420}{{\ttfamily 1412.7420}}].

\bibitem{Ball:2017nwa}
{\scshape NNPDF} collaboration, R.~D. Ball et~al., \emph{{Parton distributions
  from high-precision collider data}},
  \href{https://doi.org/10.1140/epjc/s10052-017-5199-5}{\emph{Eur. Phys. J.}
  {\bfseries C77} (2017) 663}
  [\href{https://arxiv.org/abs/1706.00428}{{\ttfamily 1706.00428}}].

\bibitem{Alekhin:2017kpj}
S.~Alekhin, J.~Bl{\"u}mlein, S.~Moch and R.~Pla\v{c}akyt{\.e}, \emph{{Parton
  distribution functions, $\alpha_s$, and heavy-quark masses for LHC Run II}},
  \href{https://doi.org/10.1103/PhysRevD.96.014011}{\emph{Phys. Rev.}
  {\bfseries D96} (2017) 014011}
  [\href{https://arxiv.org/abs/1701.05838}{{\ttfamily 1701.05838}}].

\bibitem{Accardi:2016qay}
A.~Accardi, L.~T. Brady, W.~Melnitchouk, J.~F. Owens and N.~Sato,
  \emph{{Constraints on large-$x$ parton distributions from new weak boson
  production and deep-inelastic scattering data}},
  \href{https://doi.org/10.1103/PhysRevD.93.114017}{\emph{Phys. Rev.}
  {\bfseries D93} (2016) 114017}
  [\href{https://arxiv.org/abs/1602.03154}{{\ttfamily 1602.03154}}].

\bibitem{Dulat:2015mca}
S.~Dulat, T.-J. Hou, J.~Gao, M.~Guzzi, J.~Huston, P.~Nadolsky et~al.,
  \emph{{New parton distribution functions from a global analysis of quantum
  chromodynamics}},
  \href{https://doi.org/10.1103/PhysRevD.93.033006}{\emph{Phys. Rev.}
  {\bfseries D93} (2016) 033006}
  [\href{https://arxiv.org/abs/1506.07443}{{\ttfamily 1506.07443}}].

\bibitem{Abramowicz:2015mha}
{\scshape ZEUS and H1} collaborations, H.~Abramowicz et~al., \emph{{Combination
  of measurements of inclusive deep inelastic ${e^{\pm }p}$ scattering cross
  sections and QCD analysis of HERA data}},
  \href{https://doi.org/10.1140/epjc/s10052-015-3710-4}{\emph{Eur. Phys. J.}
  {\bfseries C75} (2015) 580}
  [\href{https://arxiv.org/abs/1506.06042}{{\ttfamily 1506.06042}}].

\bibitem{Harland-Lang:2014zoa}
L.~A. Harland-Lang, A.~D. Martin, P.~Motylinski and R.~S. Thorne, \emph{{Parton
  distributions in the LHC era: MMHT 2014 PDFs}},
  \href{https://doi.org/10.1140/epjc/s10052-015-3397-6}{\emph{Eur. Phys. J.}
  {\bfseries C75} (2015) 204}
  [\href{https://arxiv.org/abs/1412.3989}{{\ttfamily 1412.3989}}].

\bibitem{Sterman:1986aj}
G.~F. Sterman, \emph{{Summation of Large Corrections to Short Distance Hadronic
  Cross-Sections}},
  \href{https://doi.org/10.1016/0550-3213(87)90258-6}{\emph{Nucl. Phys.}
  {\bfseries B281} (1987) 310}.

\bibitem{Catani:1989ne}
S.~Catani and L.~Trentadue, \emph{{Resummation of the QCD Perturbative Series
  for Hard Processes}},
  \href{https://doi.org/10.1016/0550-3213(89)90273-3}{\emph{Nucl. Phys.}
  {\bfseries B327} (1989) 323}.

\bibitem{Korchemsky:1988si}
G.~P. Korchemsky, \emph{{Asymptotics of the Altarelli-Parisi-Lipatov Evolution
  Kernels of Parton Distributions}},
  \href{https://doi.org/10.1142/S0217732389001453}{\emph{Mod. Phys. Lett.}
  {\bfseries A4} (1989) 1257}.

\bibitem{Baikov:2016tgj}
P.~A. Baikov, K.~G. Chetyrkin and J.~H. K{\"u}hn, \emph{{Five-Loop Running of
  the QCD coupling constant}},
  \href{https://doi.org/10.1103/PhysRevLett.118.082002}{\emph{Phys. Rev. Lett.}
  {\bfseries 118} (2017) 082002}
  [\href{https://arxiv.org/abs/1606.08659}{{\ttfamily 1606.08659}}].

\bibitem{RuizArriola:1998er}
E.~Ruiz~Arriola, \emph{{NLO evolution for large scale distances, positivity
  constraints and the low-energy model of the nucleon}},
  \href{https://doi.org/10.1016/S0375-9474(98)00489-8}{\emph{Nucl. Phys.}
  {\bfseries A641} (1998) 461}.

\end{thebibliography}\endgroup

\end{document}